\documentclass[11pt]{article}

\usepackage{algorithm}
\usepackage{algorithmic}
\usepackage{amsfonts}
\usepackage{amsmath}
\usepackage{amssymb}
\usepackage{amsthm}
\usepackage{bm}
\usepackage{booktabs}
\usepackage{caption}
\usepackage{graphicx}
\usepackage[utf8]{inputenc}
\usepackage{makecell}
\usepackage{mathtools}
\usepackage{multirow}
\usepackage[sort&compress]{natbib}
\usepackage{pifont}
\usepackage{soul}
\usepackage{subcaption}
\usepackage{xcolor}

\newcommand{\hashimoto}[1]{\textcolor{black}{#1}}

\title{Financial Market as a Self-Organized Ecosystem: Simulation via Learning with\\ Heterogeneous Preferences}

\author{
Ryuji Hashimoto$^{1,2}$\thanks{Corresponding author: hashimoto-ryuji419@g.ecc.u-tokyo.ac.jp},
Ryosuke Takata$^{1,2}$,
Masahiro Suzuki$^{1,2}$,\\
Yuki Tanaka$^{1,2}$,
Kiyoshi Izumi$^{1,2}$ \\
\\
$^1$ Simulacra Inc., 6-25-14 Hongo, Bunkyo-ku, Tokyo, Japan \\
$^2$ The University of Tokyo, 7-3-1 Hongo, Bunkyo-ku, Tokyo, Japan
}
\date{}

\begin{document}
\maketitle
%%%%%%%%%%%%%%% ABSTRACT %%%%%%%%%%%%%%%%%%%%%%%%%%%
% Type your abstract below
\begin{abstract}
Agent-based models provide a constructive approach to studying emergent dynamics in life-like systems composed of interacting, adaptive agents. Financial markets serve as a canonical example of such systems, where collective price dynamics arise from individual decision-making. In this modeling tradition, investor behavior has typically been captured by two distinct mechanisms---learning and heterogeneous preferences---which have been explored as separate paradigms in prior studies. However, the impact of their joint modeling on the resulting collective dynamics remains largely unexplored. We develop a multi-agent reinforcement learning framework in which agents endowed with heterogeneous risk aversion, time discounting, and information access learn trading strategies interactively within an artificial market. The experiment reveals that (i) learning under heterogeneous preferences drives agents to develop functionally differentiated strategies through interaction, rather than trait-specific rules, resulting in role specialization, and (ii) the interactions by the differentiated agents are essential for the emergence of realistic market dynamics such as fat-tailed price fluctuations and volatility clustering. Overall, this study demonstrates that the joint design of heterogeneous preferences and learning mechanisms enables the synthesis of an artificial market in which adaptive interactions drive the self-organization of a market ecology, providing a computational realization of the Adaptive Market Hypothesis.
\end{abstract}

%%%%%%%%%%%%%% KEYWORDS %%%%%%%%%%%%%%%%%%%%%%%%%%%%%%
% Please provide 5 to 6 keywords, comma-separated
% \noindent\textbf{Keywords:} Market ecology, Agent-based simulation, Multi-agent reinforcement learning, Behavioral differentiation, Self-organization, Adaptive market hypothesis

%%% ALL TEXT OF ARTICLE BELOW THIS LINE %%%
%-----------------------------------------%

\section{Introduction}
% Agent-based models: constructive modeling, not analysing the real market, but constructing the artificial market by describing investor behavior. micro behavioral pattern to macro price dynamics
\hashimoto{Agent-based models (ABMs)~\citep{abm_in_finance1,abm_in_finance2,abm_in_finance3} provide a constructive framework for studying social systems composed of interacting agents, a central theme in artificial life research~\citep{alife1,alife2}. Financial markets offer a rich testbed for such models, allowing researchers to construct artificial markets in which investor behavior is implemented as agents and to examine how collective price dynamics arise from their interactions~\citep{economy_needs_abm}.}

% Two separate mechanisms driving agent behavior: Learning and Heterogeneity
% Gap: Joint modeling is unexplored. More descriptive, 2-staged emergence
To capture investors' behavior, previous work explored two mechanisms---learning and heterogeneous preferences. Learning means that an investor discovers how to satisfy itself based on its own experiences~\citep{learning1,learning2,learning3,learning4}. Heterogeneous preferences mean differences in traits among investors~\citep{heterogeneous_preference2,heterogeneous_preference3}. Although recent studies~\citep{learning_and_heterogeneity1,learning_and_heterogeneity2,learning_and_heterogeneity3,adaptive_fcn_agent} introduced both mechanisms, the impact of such a joint modeling on the resulting dynamics remains \hashimoto{largely} unexplored. \hashimoto{When agents learn simultaneously and shape the market environment through their interactions, the relationship between preferences and behavior is no longer one-to-one. Learning-driven actions modify market states such as prices and liquidity, which in turn affect agents’ subsequent learning. As a result, the effects of preferences emerge indirectly and nonlinearly through the market environment, making the resulting behavioral differentiation and market structure difficult to predict a priori.}

% Methods
To investigate how learning and heterogeneous preferences jointly shape market structure and price dynamics, we employ a multi-agent reinforcement learning (MARL) framework. Within this framework, we design learning mechanism driven by multifaceted preferences. Agents are endowed with individual traits---including risk aversion, time discounting, and informational constraints---and approximate optimal shared-policy\footnote{\hashimoto{Here, the term {\em policy} follows standard reinforcement learning terminology~\citep{rl} and refers to a parametric mapping from an agent’s observations to action probabilities, rather than to any notion of economic or institutional policy.}
} in a market environment through repeated interactions with one another.

% experiment: investigating RQs
In our experiment, we conduct simulations with the \hashimoto{heterogeneous learning} agents to investigate (i) behavioral differentiation arising from interactive learning within a unified framework where such heterogeneity in preferences exists and (ii) the emergence of price dynamics through the interactions of these agents. 
For (i), \hashimoto{first, within a shared-policy setting, we examine how agents’ traits are expressed through learned order decisions. Rather than assuming a direct trait–action mapping, we show that interactive learning with heterogeneous preferences generates heterogeneous behavior, as evidenced both in observable actions and in internal policy representations. Crucially, the resulting patterns of specialization depend on the prior distribution of agent traits: even with identical support, different initial trait distributions lead to qualitatively distinct differentiation structures, indicating that behavior emerges through interaction-mediated learning rather than trait conditioning alone. An ablation study removing preference heterogeneity further shows that, without heterogeneous preferences, agents fail to develop comparable specialization and achieve significantly lower aggregate utility, highlighting the meso-level role of learning-driven differentiation in forming complementary trading roles.}
For (ii), we show that our model achieves greater realism than baseline models that incorporate either preference heterogeneity or learning, based on several metrics. By allowing heterogeneous agents to adapt their behavior through learning under endogenous market feedback, key empirical regularities of financial markets emerge---such as fat-tailed return distributions and volatility clustering---that characterize the complex price dynamics observed in reality. \hashimoto{Taken together, we show that through learning-driven behavioral differentiation among agents with heterogeneous preferences under market interaction, market niches become endogenously structured, and a population self-organizes into an ecosystem of interdependent trading roles that provides the foundation for the emergence of complex price dynamics. This perspective aligns closely with the Adaptive Market Hypothesis (AMH)}~\citep{amh1,amh2}\hashimoto{, which views financial markets as evolving ecosystems. Our framework provides a mechanistic account of how such ecosystems self-organize through learning and interaction, giving rise to observed price dynamics.}

\section{Related Work}
% 投資家の異質性が，金融市場のstylized factsや市場アノマリー，クラッシュなどの極端事象を再現する上で重要であるとの認識から，ABMによるアプローチが発達
% リスク回避度 -> FCN，
% time horizon -> ...
% (+ heterogeneous beliefを加える，．普通はchartistとfundamentalのweighting)
% 一方，市場のnonequibrium natureを再現するために学習メカニズムの設計も注目
The recognition that investor heterogeneity is crucial for reproducing stylized facts, market anomalies, and extreme events in financial markets has driven the development of ABM approaches~\citep{heterogeneous_preference2,heterogeneous_preference3}. For example, heterogeneity of risk aversions is proved to \hashimoto{have an effect} on equity risk premium~\citep{risk_aversion_term_to_risk_premium} and excess volatility~\citep{risk_aversion_term_to_excess_volatility1,risk_aversion_term_to_excess_volatility2}. \citet{dynamic_risk_aversion} demonstrate that heterogeneous and dynamic risk aversion of investors plays a role to generate volatility clustering. Also, heterogeneous time horizon is demonstrated to cause information delay, resulting in persistence of stock price fluctuations~\citep{time_horizon}.

To capture the non-equilibrium nature of financial markets, the design of learning \hashimoto{mechanisms have} also attracted attention. A representative example is the adaptive belief system~\citep{learning2,learning3,adaptive_belief_system3}, a framework in which agents iteratively update their forecasting rules based on past performance, thereby allowing market expectations to evolve endogenously. Other approaches, such as classifier systems~\citep{learning4} and genetic algorithms~\citep{abm_in_finance1}, have likewise been employed to model how agents adapt to their environment and to explain the emergence of abrupt phenomena such as bubbles and crashes.

Although recent studies have incorporated both preference heterogeneity and learning into their models~\citep{learning_and_heterogeneity1,learning_and_heterogeneity2,learning_and_heterogeneity3,adaptive_fcn_agent}, \hashimoto{these two elements are typically introduced at separate modeling layers rather than being integrated within a single learning process. In many such approaches, preference heterogeneity is reflected through differences in predefined behavioral rules, and learning mechanisms are typically applied over this predefined strategy space. While such designs are effective for studying strategy competition, they make it difficult to examine how preferences are translated into behavior through learning itself.}

\hashimoto{We argue that learning and preference heterogeneity are the fundamental ingredients that give rise to financial markets as ecosystems. In contrast to the Efficient Market Hypothesis, which seeks to understand market fluctuations through the lens of physical dynamics, the AMH~\citep{amh1,amh2} conceptualizes financial markets as ecosystems in which heterogeneous participants coevolve through learning and interaction. Empirical support for the AMH has grown~\citep{empirical_amh_survey,empirical_amh_predictability1,empirical_amh_predictability2} through studies that observe temporal variation in market efficiency and return predictability under changing market conditions. Within the AMH framework, several ABMs incorporate heterogeneity and learning primarily through switching among a finite set of strategies~\citep{abm_amh2,abm_amh3,abm_amh1}. These models have successfully shown that interactions among heterogeneous strategies can generate realistic price dynamics and time-varying market efficiency. However, they typically take the existence of distinct strategic roles as given, and therefore do not explain how such ecosystem-like structures arise in the first place. Models that explicitly examine the internal generative mechanisms through which heterogeneity and learning interact to produce these ecological patterns remain lacking. This study addresses this gap by constructing an ABM in which individual agents’ learning is driven by heterogeneous preferences, allowing us to demonstrate how learning dynamics and endogenous market feedback jointly give rise to ecosystem-like self-organization.
}

%We argue that learning and preference heterogeneity are the key drivers for establishing a multi-level emergent view of financial markets. In biology, complex living systems are explained through hierarchical emergent mechanisms~\citep{collective_intelligence}, and similar multi-scale order is suggested to exist in human societies as another form of complex system~\citep{complexity_and_economy,complex_adaptive_system,future_economics}. However, a concrete explanation of such mechanisms in financial markets has not yet been attempted.

\section{Method}

\begin{figure}[tbp]
  \centering
  \includegraphics[width=0.7\linewidth]{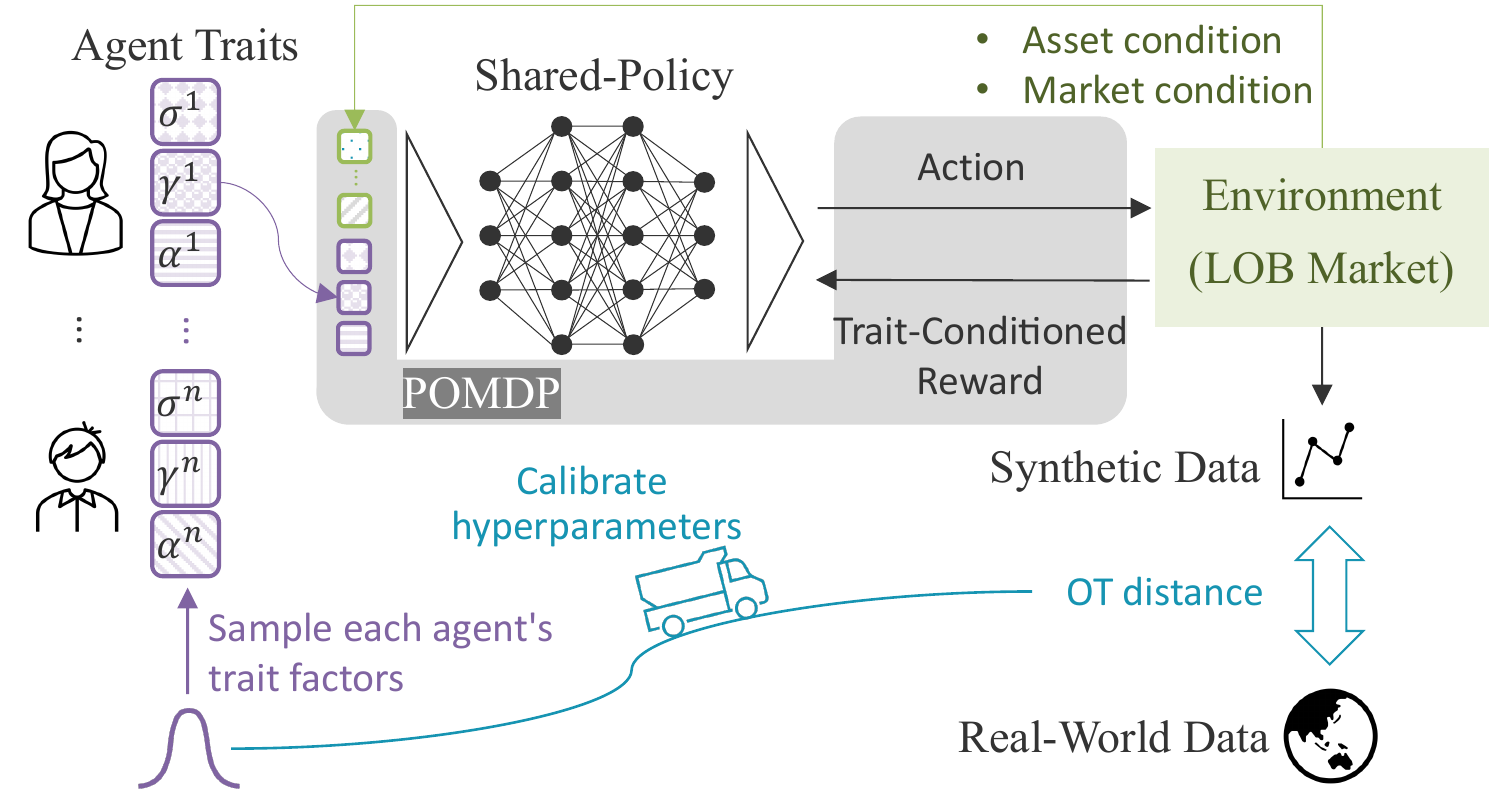}
  \caption{Conceptual diagram of our MARL-based ABM for financial market simulations. Each agent is assigned individual traits when the simulation starts. A shared-policy, learned to satisfy each agent's specific preference, governs agent behavior within a \hashimoto{limit order book (LOB)} market environment. The prior distribution of the agents' trait factors are calibrated using \hashimoto{optimal transport (OT)} so that the synthetic price series aligns with real data.}
  \label{Fig:conceptual_diagram}
\end{figure}

Figure~\ref{Fig:conceptual_diagram} describes our simulation methodology. The core components of the method are a formulation as a partially observable Markov decision process (POMDP) in which agent traits are embedded in both the observations and the rewards, combined with shared-policy learning~\citep{shared_policy_learning,learning_and_heterogeneity3}, and a calibration of the distribution of agent traits via optimal transport (OT)~\citep{ot_based_simulation_evaluation}.
\begin{enumerate}
\item \textbf{POMDP formulation}: Each agent operates under a POMDP, where its observation vector includes trait factors, allowing the policy to condition on {\em who it is}. These traits are also reflected in the reward function, such that each agent is pursuing a different goal, encouraging the emergence of preference-driven behavioral diversity.
\item \textbf{Shared-policy learning across heterogeneous agents}: A single neural network policy is trained to map observations into actions. Heterogeneous traits modify each agent’s observation, reward, and state–visitation distribution, so that a family of conditional policies emerge within a single neural network.
\item \textbf{Data-driven calibration via OT}: The prior distribution of agent traits is calibrated at the population level to minimize the OT distance between the synthetic data generated by the simulation and real-world financial data, thereby identifying plausible regions of parameter space~\citep{calibration1,calibration2}.
\end{enumerate}
This section describes the method in the following order: (1) the limit order book (LOB) market environment where agents interact; (2) the POMDP formulation defined by the observation, action, and reward structure; (3) the learning procedure of the shared-policy; and (4) the OT-based calibration of agent trait distributions.

\subsection{Simulation Structure}

\begin{figure}[tbp]
  \centering
  \includegraphics[width=0.6\linewidth]{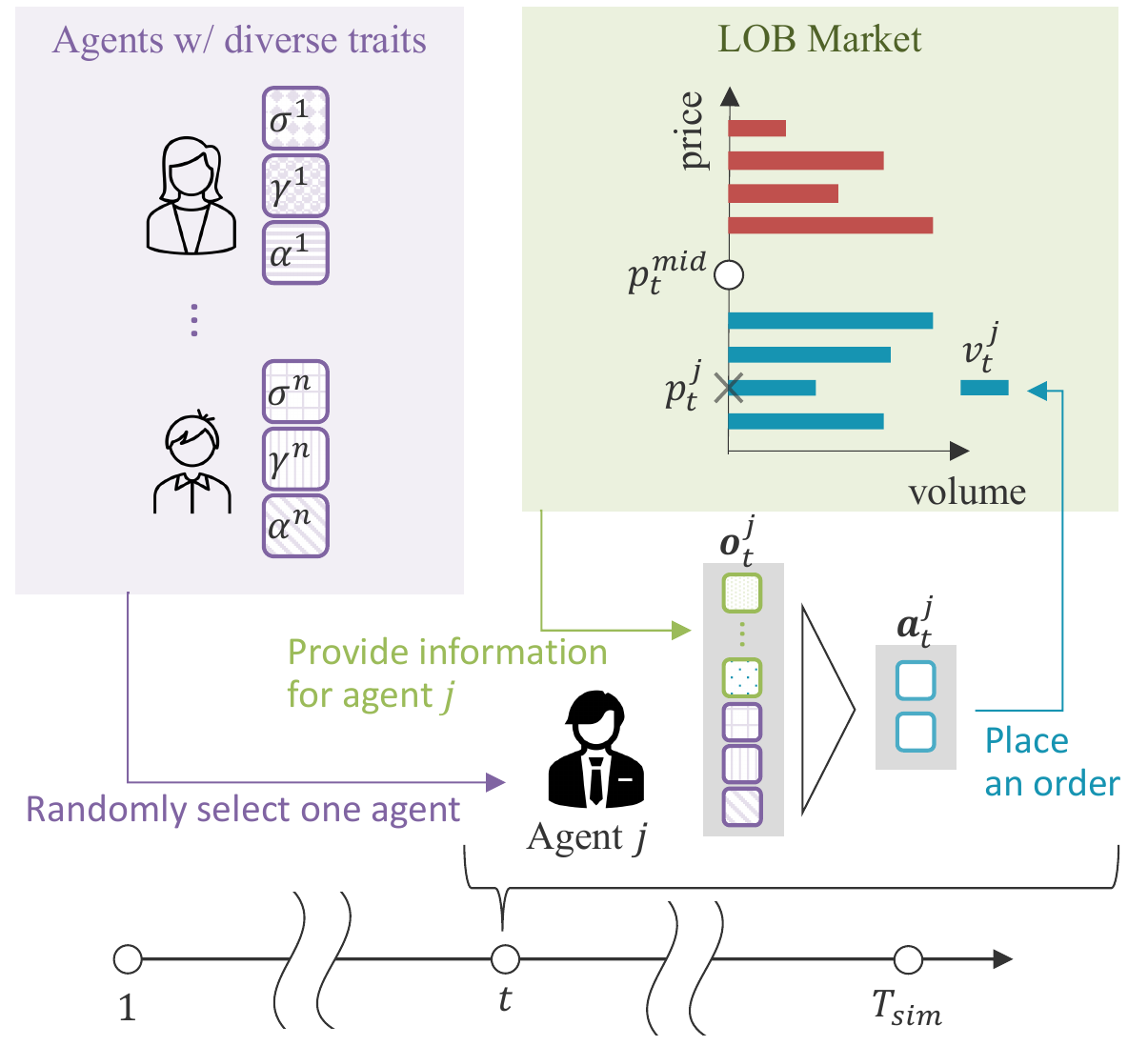}
  \caption{Structure of the simulation. At each time step $t$, one of the $n$ agents is selected to submit an order. Their order is placed into a LOB market, which matches buy and sell orders to determine transactions and update the market state.}
  \label{Fig:simulation_structure}
\end{figure}

Assume $n\in\mathbb{N}$ agents operate in a single market setting, with each simulation consisting of $T_{sim}\in\mathbb{N}$ time steps. Figure~\ref{Fig:simulation_structure} illustrates the structure of our simulation settings. At each time step $t \in \{1,\ldots,T_{sim}\}$, a randomly selected agent $j \in \{1,\ldots,n\}$ is allowed to place an order. Agent $j$ submits an order specifying a signed order volume $v_t^j \in \mathbb{Z}$ and an order price $p_t^j \in \mathbb{R}_+$. The sign of $v_t^j$ indicates whether the agent intends to buy or sell\footnote{For instance, if $v_t^j = -2$ and $p_t^j = 300.0$, agent $j$ submits an order \hashimoto{at time $t$} to sell $2$ units of the stock at a price of 300.0.}.

The market operates under the double auction rule, where orders are collected into a central LOB. The LOB displays anonymized prices and volumes at which agents are willing to buy or sell the stock, and the market matches orders.

\subsection{POMDP and Shared-Policy Learning}
Let $\mathcal{T}^j$ denote the set of time steps at which agent $j$ is selected to place an order, with components $t_1^j < t_2^j < \ldots < t_{\iota_j}^j \in \mathcal{T}^j$, where $\iota_j$ represents the number of times agent $j$ is selected during the simulation. % satisfying the condition $\sum_j \iota_j = T_{sim}$. 
Agent $j$ aims to learn the optimal policy $\pi^{j*}$ that maximizes its expected cumulative discounted reward, defined as:
{\small
\begin{equation}
    \begin{split}
        \pi^{j*}=\arg\max_{\pi^j}\mathbb{E}_{\pi^j}\left[\sum_{i=1}^{\iota_j}(\gamma^j)^ig(\bm{o}_{t_i^j}^j,\bm{a}_{t_i^j}^j;j)\right]
    \end{split}
\end{equation}}
where $\mathbb{E}_{\pi^j}[\cdot]$ denotes the expectation under the policy $\pi^j$, and $\gamma^j$ is agent $j$'s discount factor. The reward function $g: \mathbb{R}^{11} \times \mathbb{R}^2 \to \mathbb{R}$ depends on the agent's observation $\bm{o}_{t_i^j}^j \in \mathbb{R}^{11}$ and action $\bm{a}_{t_i^j}^j \in \mathbb{R}^2$ at time step $t_i^j$.

\subsubsection{Observation}

Let $t = t_i^j$ in the following sections. The observation vector $\bm{o}_t^j \in \mathbb{R}^{11}$ is a numerical representation of the market conditions and agent $j$'s personal trait factors. It comprises the following eleven components:
\begin{itemize}
\item Holding asset ratio: $\frac{w_t^j p_t^{mid}}{x_t^j}$
\item Holding asset-to-maximum order volume ratio: $\frac{w_t^j}{v_{max}}$
\item Inverted buying power: $\frac{p_t^{mid}}{c_t^j}$
\item Return: $r_{[t_{i-1}^j, t_i^j]}$
\item Volatility: $V_{[t_{i-1}^j, t_i^j]}$
\item Asset volume-to-existing buy (sell) order volumes ratio: $\frac{|w_t^j|}{b_t^\xi},~\frac{|w_t^j|}{s_t^\xi}$
\item Blurred fundamental return: $\tilde{r}_t^{f,j}$
\item Uninformedness: $\sigma^j$
\item Risk aversion term: $\alpha^j$
\item Discount factor: $\gamma^j$
\end{itemize}
$w_t^j \in \mathbb{Z}$ and $c_t^j \in \mathbb{R}$ represent agent $j$'s stock position and holding cash amount at time step $t$, respectively. At the beginning of the simulation, $w_0^j$ and $c_0^j$ are randomly set as $w_0^j\sim Ex(w)$, $c_0^j\sim Ex(c)$ for each $j$, where $Ex(\lambda)$ indicates an exponential distribution with expected value $\lambda$, and $w$ and $c$ are the expected values of initial asset volume and cash amount. The agent's total asset value, $x_t^j$, is given by $x_t^j = c_t^j + w_t^j p_t^{mid}$, where $p_t^{mid}$ denotes the mid price at time $t$\footnote{The mid price is calculated as the average of the best bid and ask prices. The best bid (ask) price refers to the highest (lowest) price buyers (sellers) are willing to trade on the LOB.}. $v_{max} \in \mathbb{N}$ is an exogenously determined constant that specifies the maximum order volume an agent can place at time step $t$. $r_{[t_{i-1}^j,t_i^j]} \in \mathbb{R}$ represents the logarithmic return of the price series between $t_{i-1}^j$ and $t_i^j$, calculated as follows:
{\small
\begin{equation}
    \begin{split}
    r_{[t_{i-1}^j,t_i^j]}=\frac{1}{t_i^j-t_{i-1}^j}\sum_{t'=t_{i-1}^j+1}^{t_i^j}\log\frac{p_{t'}^{mid}}{p_{{t'}-1}^{mid}}
    \end{split}
\end{equation}}
$V_{[t_{i-1}^j, t_i^j]}\in\mathbb{R}_+$ denotes the volatility:
{\small
\begin{equation}
    \begin{split}
    V_{[t_{i-1}^j, t_i^j]}=\frac{1}{t_i^j-t_{i-1}^j}\sum_{t'=t_{i-1}^j+1}^{t_i^j}\left(\log\frac{p_{t'}^{mid}}{p_{t'-1}^{mid}}-r_{[t_{i-1}^j,t_i^j]}\right)^2
    \end{split}
\end{equation}}
$b_t^\xi,s_t^\xi\in\mathbb{R}_+$ denote the weighted sum of order volumes within the price range on the LOB, extending up to $\xi p_t^{mid}$ away from the mid price $p_t^{mid}$
{\small
\begin{eqnarray}
b_t^\xi &=&\sum_{p_t^{mid}(1-\xi)<p<p_t^{mid}}v_t^b(p)\exp\left(-\omega^b\frac{|p_t^{mid}-p|}{p_t^{mid}}\right)\label{Eq:b_t_xi}\\
s_t^\xi &=&\sum_{p_t^{mid}<p<p_t^{mid}(1+\xi)}v_t^s(p)\exp\left(-\omega^s\frac{|p_t^{mid}-p|}{p_t^{mid}}\right)\label{Eq:s_t_xi}
\end{eqnarray}}
$\xi\in\mathbb{R}_+$ is a parameter that determines the maximum depth. $v_t^b(\cdot)$ and $v_t^s(\cdot)$ represent the buy and sell order volumes at specific prices displayed in the LOB. As shown in Equations~(\ref{Eq:b_t_xi}) and (\ref{Eq:s_t_xi}), the existing order volumes are weighted by their distances from the mid price. $\omega^b, \omega^s \in \mathbb{R}_+$ are parameters that control the decay rate. The agent $j$ receives a fundamental return blurred by Gaussian noise, the scale of which depends on their level of uninformedness $\sigma^j \in \mathbb{R}_+$.
{\small
\begin{equation}
    \begin{split}
    \tilde{r}_t^{f,j}=\log\frac{p_t^f}{p_t^{mid}}+\eta_t^j,~~\eta_t^j\sim\mathcal{N}(0,(\sigma^j)^2)\label{Eq:blurred_fundamental_return}
    \end{split}
\end{equation}}
$p_t^f$ represents the actual fundamental price of the stock. In real-world stock markets, investors estimate a company's intrinsic value, known as the fundamental price, based on various publicly available information and make trading decisions accordingly. An uninformedness $\sigma^j$ is introduced in Equation~(\ref{Eq:blurred_fundamental_return}) to capture differences in investors' ability to accurately perceive the fundamental price. The term $\mathcal{N}(0, (\sigma^j)^2)$ denotes a Gaussian distribution with a mean of $0$ and a variance of $(\sigma^j)^2$.

The agent $j$'s trait factors $\sigma^j$, $\alpha^j$, and $\gamma^j$ are randomly determined before each simulation as $\forall j~~\sigma^j\sim\mathcal{N}\left(\mu^\sigma,(\lambda^\sigma)^2\right),~\alpha^j\sim\mathcal{N}\left(\mu^\alpha,(\lambda^\alpha)^2\right),~\gamma^j\sim U(\lambda^\gamma,\gamma^{max})$ where $U(\lambda^\gamma,\gamma^{max})$ indicates uniform distribution with minimum and maximum values $\lambda^\gamma$ and $\gamma^{max}$.

\hashimoto{In summary, each agent’s observation vector includes not only the agent traits (uninformedness, risk aversion, and the discount factor) but also variables describing the agent’s own asset position (holding asset ratio, the holding asset-to-maximum order volume ratio, and inverted buying power), and market-level indicators (returns, volatility, the ratio of the agent’s asset volume to existing buy and sell order volumes, and the fundamental return). These variables are standard observation components widely adopted in agent-based finance models~\citep{abm_rl1,abm_rl2}. While the choice of observation variables follows established practice, our implementation incorporates an additional design consideration: to enable stable learning under a shared-policy framework across heterogeneous agents, all observation components are integrated into the input vector in a normalized form.}

\subsubsection{Action}
The action $\bm{a}_t^j\in\mathbb{R}^{2}$ consists of the following two components to form the agent $j$'s order.
\begin{itemize}
\item Scaled order volume $\tilde{v}_t^j$
\item Scaled order margin $\tilde{r}_t^j$
\end{itemize}
Using $\tilde{v}_t^j,\tilde{r}_t^j\in[-1,1]$, the signed order volume $v_t^j$ and order price $p_t^j$ is decided as follows.
{\small
\begin{equation}
    \begin{split}
    v_t^j=\lceil v_{max}\tilde{v}_t^j\rceil,~~ p_t^j=p_t^{mid}-r_{max}\cdot\mathrm{sign}(\tilde{v}_t^j)\cdot\tilde{r}_t^j\cdot p_t^{mid}\label{Eq:action2order}
    \end{split}
\end{equation}}
Here, $\lceil\cdot\rceil$ is a ceiling function. $r_{max}\in\mathbb{R}_+$ denotes the maximum scaled order price margin.

\subsubsection{Reward}

The immediate reward for agent $j$ at time step $t$ is defined as $r_t^j = g(\bm{o}_t^j, \bm{a}_t^j; j)$, where the reward function $g(\cdot, \cdot)$ is specified as follows.
\begin{equation}
g(\bm{o}_t^j,\bm{a}_t^j;j)=u_t^j-\beta^{short}\textbf{1}(w_t^j<0)-\beta^{cash}\textbf{1}(c_t^j<0)-\beta^{illiquidity} l_t-\beta^{fundamental}R_t^f\label{Eq:reward}
\end{equation}
where, $\beta^{short}, \beta^{cash}, \beta^{illiquidity}, \beta^{fundamental} \in \mathbb{R}_+$ are weights corresponding to each component of the reward function. The components of the reward function (Equation~(\ref{Eq:reward})) can be categorized into terms related to individual learning (the first, second, and third terms) and those related to collective learning (the fourth and fifth terms). The first term, $u_t^j \in [-1,1]$, represents agent $j$’s utility: 
\begin{align}
u_t^j=\frac{2}{\pi}\times\mathrm{Arctan}\left(\omega^u(x_t^j+r_{[t_{i-1}^j,t_i^j]}w_t^jp_t^{mid}-\frac{\alpha^j}{2}V_{[t_{i-1}^j, t_i^j]}|w_t^jp_t^{mid}|)\right)\label{Eq:utility}
\end{align}
where $\omega^u$ is a scaling factor for the utility. Equation~(\ref{Eq:utility}) models the agent $j$’s constant absolute risk averse (CARA) utility~\citep{fcn2}, capturing the decrease in utility due to volatility and the agent’s risk aversion parameter. The $\mathrm{Arctan}$ function compresses utility values to account for varying wealth scales. The second and third terms in Equation~(\ref{Eq:reward}) impose penalties for short selling and cash shortages, respectively, where $\textbf{1}(\cdot)$ denotes an indicator function. The fourth term represents an illiquidity penalty, reflecting investor discomfort when the stock market exhibits illiquidity. Illiquidity, $l_t$, is calculated as follows: 
\begin{equation}
    l_t=\frac{1}{b_t^\xi}+\frac{1}{s_t^\xi}+\omega^l\left(\frac{\max(b_t^\xi,s_t^\xi)}{\min(b_t^\xi,s_t^\xi)}-1\right)\label{Eq:illiquidity}
\end{equation}
where $l_t$ increases when either the buy or sell order volume in the LOB is low or when the order volumes are imbalanced. Here, $\omega^l$ is a scaling factor for the order volume imbalance. The fifth term in Equation~(\ref{Eq:reward}) is a penalty for deviation from the fundamental price. The integrated fundamental return, $R_t^f$, is calculated as follows:
\begin{align}
\begin{split}
\MoveEqLeft R_t^f = \sum_{\tau < t' \leq t} \left|\frac{1}{t+1-t'} \log \frac{p_{t'}^f}{p_{t'}^{mid}} \right| \\
\MoveEqLeft \text{s.t.}~ \mathrm{sign}\left( \log \frac{p_{\tau}^f}{p_{\tau}^{mid}} \right) \neq \mathrm{sign}\left( \log \frac{p_t^f}{p_t^{mid}} \right),~~~ \mathrm{sign}\left( \log \frac{p_{t'}^f}{p_{t'}^{mid}} \right) = \mathrm{sign}\left( \log \frac{p_t^f}{p_t^{mid}} \right),~ \forall \tau < t' < t
\end{split}
\end{align}

% {\small
% \begin{align}
% \begin{split}
% \MoveEqLeft R_t^f=\sum_{\tau<t'\leq t}\left|\log\frac{p_{t'}^f}{p_{t'}^{mid}}\right|\\
% \MoveEqLeft s.t.~~\mathrm{sign}(\log\frac{p_{\tau}^f}{p_\tau^{mid}})\neq\mathrm{sign}(\log\frac{p_{t}^f}{p_t^{mid}}),\\
% \MoveEqLeft~~~~~~~~~ \forall\tau<t'~\mathrm{sign}(\log\frac{p_{t'}^f}{p_{t'}^{mid}})=\mathrm{sign}(\log\frac{p_{t}^f}{p_t^{mid}})
% \end{split}
% \end{align}}
The term $\beta^{fundamental} R_t^f$ imposes a larger penalty as the market price deviates from the fundamental price in the same direction for an extended period. This penalty \hashimoto{encourages} agents to collectively learn the consensus that market prices tend to fluctuate around the fundamental price.

Rather than hard-coding behavioral rules, we design reward functions that encode individual preferences. It specifies {\em what to optimize}, not {\em how to act}, providing a meta-level guidance for learning, without prescribing fixed action rules.

\begin{algorithm}[tbp]
    \caption{The shared-policy framework for learning heterogeneous trading strategies in financial market simulations.}
    \label{Alg:learn_heterogeneous_marl}
    \begin{algorithmic}[1]
    \REQUIRE %$n$, the number of agents, $T_{sim}$, the number of total time steps in a simulation, $w$ and $c$, the expected values of initial asset volume and cash amount, $\lambda^\sigma$, $\lambda^{\gamma}$, and $\lambda^{\alpha}$, the parameters controlling the degree of heterogeneity in the agents' trait factors, $\mu^\sigma$, $\gamma^{max}$, and $\mu^{\alpha}$, the additional parameters for the prior distribution of the trait factors, $\omega^b$, $\omega^s$, $\omega^u$, and $\omega^l$ the scaling factors, $v_{max}$ and $r_{max}$, the maximum order volume and the scaled margin from the current mid price, $\beta^{short}$, $\beta^{cash}$, $\beta^{illiquidity}$, and $\beta^{fundamental}$, the weights of the reward function on the penalties, 
    $T_{rollout}$, the rollout length, $\beta^{actor}$ and $\beta^{critic}$, the learning rate for the actor and critic networks, %$\epsilon$, the clipping parameter for the surrogate objective of PPO, $\lambda$, discount factor for GAE, 
    and orthogonally initialized $\bm{\theta}$ and $\bm{\phi}$, the parameters for actor and critic networks, $L_{actor}(\cdot)$ and $L_{critic}(\cdot)$, loss functions for updating the networks. %$\pi_{\bm{\theta}}$ and $\bm{\phi}$, the parameters for critic network $C_{\bm{\phi}}$, $L_{actor}(\cdot)$ and $L_{critic}(\cdot)$, loss functions for updating actor and critic networks.
    \STATE Set rollout buffer for all agents empty $\forall j~ D_j=\{\}$
    \WHILE{not converge}
        \STATE Set trait factors of all agents by randomly sample $\forall j~w_0^j,c_0^j,\sigma^j,\gamma^j$, and $\alpha^j$ from prior distributions.%$\forall j~w_0^j\sim Ex(w)$, $c_0^j\sim Ex(c)$, $\sigma^j\sim\mathcal{N}(\mu^\sigma,\lambda^\sigma)$, $\gamma^j\sim U(\lambda^\gamma, \gamma^{max})$, and $\alpha^j\sim\mathcal{N}(\mu^\alpha,\lambda^\alpha)$
        \STATE Set previous order times of all agents $\forall j~ t_0^j\leftarrow0$
        \FOR{$t\in \{1,\ldots,T_{sim}\}$}
            \STATE Update fundamental price $p_t^f$
            \STATE Randomly sample an agent $j\in\{1,\ldots, n\}$
            \STATE Retrieve the last order time of agent $j$, $t_{i-1}^j$ and set $t_i^j \leftarrow t$ (current time step, $i\in\{1,\ldots\iota_j\}$)
            \STATE Get observation for the agent $j$, $\bm{o}_{t_i^j}^j$
            \STATE Sample action $\bm{a}_{t_i^j}^j$ and its probability {\small $\pi_{\bm{\theta}_{t_i^j}}(\bm{a}_{t_i^j}^j | \bm{o}_{t_i^j}^j)$}
            \STATE Convert action $\bm{a}_{t_i^j}^j$ to an order, submit the order to the market, and the market execute orders
            \STATE Calculate reward for agent $j$, $r_{t_i^j}^j = g(\bm{o}_{t_i^j}^j, \bm{a}_{t_i^j}^j; j)$
            \IF{$1<i$}
                \STATE {\small $D_j=D_j~\cup~\left[\bm{o}_{t_{i-1}^j}^j,\bm{a}_{t_{i-1}^j}^j,\log \pi_{\bm{\theta}_{t_{i-1}^j}}, r_{t_i^j}^j,\bm{o}_{t_i^j}^j\right]$}
                \IF{length of $D_j$ equals $T_{rollout}$}
                    \STATE Update actor and critic using rollout {\small $D_j$, $\bm{\theta}_t\leftarrow\bm{\theta}_{t-1}-\beta^{actor}\nabla L_{actor}(\bm{\theta}_{t-1})$, $\bm{\phi}_t\leftarrow \bm{\phi}_{t-1}-\beta^{critic}\nabla L_{critic}(\bm{\phi}_{t-1})$}
                    \STATE Clear rollout buffer for agent $j$ $D_j=\{\}$
                \ENDIF
            \ENDIF
        \ENDFOR
    \ENDWHILE
    \end{algorithmic}
\end{algorithm}

\subsubsection{Shared-Policy Learning}

\hashimoto{To study the role of preference heterogeneity under a learning mechanism, we adopt a shared-policy learning framework in which all agents update a single policy $\pi_{\bm{\theta}}$. This design keeps the learning architecture and optimization process identical across agents, allowing behavioral differences to be attributed to heterogeneous preferences and interaction effects. Policy optimization is performed using proximal policy optimization (PPO)~\citep{ppo}, as summarized in Algorithm~\ref{Alg:learn_heterogeneous_marl}.}

\hashimoto{Following standard practice in on-policy reinforcement learning~\citep{gae}, each agent maintains its own experience buffer $D_j$, and policy updates are carried out only after collecting a fixed rollout length $T_{\mathrm{rollout}}$ from each buffer. This implementation ensures that updates are based on temporally coherent trajectories generated by individual agents, which improves training stability in a non-stationary multi-agent market environment.}

%We train a single shared-policy across all agents using proximal policy optimization (PPO)~\citep{ppo}. The details of the learning procedure are outlined in Algorithm~\ref{Alg:learn_heterogeneous_marl}. To preserve individual-level variation and stabilize learning, each agent maintains its own experience buffer $D_j$, and the policy $\pi_{\bm{\theta}}$ is updated only when each buffer reaches a fixed rollout length $T_{rollout}$. This design is aimed to prevent the mixing of trajectories generated under different agent preferences, which could otherwise dilute the distinct behavioral strategies shaped by their respective traits.

%To capture heterogeneous behaviors while maintaining scalability, we train a single shared-policy across all agents using proximal policy optimization (PPO)~\citep{ppo}. The details of the learning procedure are outlined in Algorithm~\ref{Alg:learn_heterogeneous_marl}. To preserve individual-level behavioral variation and ensure learning stability, each agent maintains its own experience buffer $\forall j~~D_j$, and the shared-policy $\pi_{\bm{\theta}}$ is updated only when the buffer reaches a fixed rollout length $T_{rollout}$. This design is aimed to prevent the mixing of semantically heterogeneous trajectories and allow the reinforcement signal to remain consistent with each agent’s trait-driven preferences, facilitating stable and interpretable policy learning across a diverse agent population.

\subsection{Calibration of Agent Trait Distribution}

Since agent traits shape market-level behavior, it is crucial to calibrate their population-level distribution such that the simulated data resembles empirical data. To achieve this, we adopt the OT-based method proposed by \citet{ot_based_simulation_evaluation}, which quantitatively compares the simulation outputs and empirical data. We begin by defining points $\bm{x}_k\in\mathbb{R}^d,~ k\in\{1,\ldots,K\}$, where each $\bm{x}_k$ is a $d$-dimensional feature vector sampled from stock price series data. \hashimoto{In the experiment, we use three types of feature representations for calibration, as detailed below.} These points form a point cloud, represented as $X={}^\top\!\begin{pmatrix}\bm{x}_1 & \ldots & \bm{x}_K\end{pmatrix}\in\mathbb{R}^{K\times d}$, where $K$ denotes the number of points. The OT distance between the two point clouds $X_{syn}\in\mathbb{R}^{K\times d}$ and $X_{real}\in\mathbb{R}^{L\times d}$, is defined as:
{\small
\begin{align}
\MoveEqLeft OT(X_{\mathrm{syn}}, X_{\mathrm{real}}) = \min_{P \in \mathbb{R}^{K \times L}}\sum_{k=1}^K \sum_{l=1}^L \left\| X_{\mathrm{real}}^k - X_{\mathrm{syn}}^l \right\|_2^2 P^{k,l}\nonumber \\
\text{s.t.}~~ & \forall k, l~~ 0 \leq P^{k,l}, ~~ \sum_{k=1}^K P^{k,l} = \tfrac{1}{L}, ~~ \sum_{l=1}^L P^{k,l} = \tfrac{1}{K}
\end{align}}
where $X_{syn}$ and $X_{real}$ are the point clouds derived from synthetic and real data, respectively. We denote the $k$-th row element of the point cloud $X$ as $X^k$. $P^{k,l}$ denotes the $k$-th row and $l$-th column element of the matrix $P\in\mathbb{R}_{\ge0}^{K\times L}$. \hashimoto{The OT distance defined above corresponds to the squared 2-Wasserstein distance between two empirical distributions represented as point clouds.} We define three types of points, $\bm{x}^r, \bm{x}^t\in\mathbb{R}^{1}$, and $\bm{x}^{as}\in\mathbb{R}^9$ to see the underlying order of return distribution, tail return distribution, and absolute return time series in the synthetic data. Let us define a series of one-minute log-return data, arranged in chronological order at equal intervals, and standardized to have a sample mean of $0$ and a sample standard deviation of $1$. Let this standardized series be denoted as $\left((r_{t,i})_{t=1}^{T_{len}}\right)_{i=1}^{N_{days}}$, where $T_{len}$ represents the length of each time series, and $N_{days}$ denotes the total number of collected series. Using $r_{t,i}$, the points are represented as:
{\small
\begin{align}
    \begin{split}
    \MoveEqLeft\bm{x}^r=\begin{pmatrix}r_{t,i}\end{pmatrix},~~~\bm{x}^t=\begin{pmatrix}\tilde{r}_{(k)}\end{pmatrix},\\
    \MoveEqLeft\bm{x}^{as}={}^\top\!\begin{pmatrix}|r_{t,i}| & |r_{t+1,i}| & |r_{t+10,i}| & \ldots & |r_{t+70,i}|\end{pmatrix}
    \end{split}
\end{align}}
where $\tilde{r}_{(k)}$ is calculated as follows. Let $r_{(N)},\ldots,r_{(1)}$ denote the descending order statistics of samples of absolute log-returns $|r_1|,\ldots,|r_N|$ of size $N~(=T_{len}N_{days})$,
{\small
\begin{eqnarray}
\tilde{r}_{(k)}=\log \frac{r_{(N-k+1)}}{r_{(N-K)}},~ k\in\{1,\ldots,K\}
\end{eqnarray}}
$\tilde{r}_{(k)}$ was introduced to see the underlying order of the tail distribution of log-returns, drawing from \citet{hill}. We define the OT distances between real and synthetic point clouds computed using the points $\bm{x}^r$, $\bm{x}^t$, and $\bm{x}^{as}$ as $OT^r$, $OT^t$, and $OT^{as}$, respectively. During calibration, the agents with each candidate combination of $\lambda^{\sigma}$, $\lambda^{\alpha}$, and $\lambda^{\gamma}$ are independently trained, and compared with weighted average of OT distances $\bar{OT}=\omega^rOT^r+\omega^tOT^t+\omega^{as}OT^{as}$ 
where $\omega^r,\omega^t,\omega^{as}\in\mathbb{R}_+$ are the weights assigned to each distance.

\section{Experiment}
Through the experiment, we conduct market simulations using agents trained by our method, and investigate the following two research questions: \hashimoto{\textbf{RQ1}: How do heterogeneous preferences, when coupled with adaptive learning and interaction, give rise to a structured market ecology characterized by differentiated trading roles? 
\textbf{RQ2}: How does the resulting market ecology shape the characteristic dynamics of stock prices?
To address RQ1, we analyze how agent traits are reflected in the organization of behavioral patterns that emerge through interaction and learning, focusing on the formation of differentiated roles within the market. To address RQ2, we examine how the self-organized market ecology gives rise to aggregate price dynamics by comparing the simulated outcomes with baseline models in which either learning or preference heterogeneity is absent. Together, these analyses allow us to elucidate the mechanisms through which learning, heterogeneous preferences, and interaction jointly lead to realistic stock market dynamics.}

In the experiment, we searched the hyperparameters $\lambda^\sigma$, $\lambda^\alpha$, and $\lambda^\gamma$ using OT-based calibration method. For real data, we used FLEX-FULL historical tick data~\citep{flex_full}. Details of the hyperparameters and data descriptions are provided in the Appendix~\ref{App:pomdp}, \ref{App:training}, and \ref{App:data}.

\subsection{Baselines}
We compare the realism of the simulation results between our method and the following baseline ABMs.
\begin{itemize}
\item \textbf{Zero intelligence (ZI-) Agent}~\citep{zero_intelligence}: In financial market simulations, ZI-Agents serve as a behaviorally naive benchmark. They submit one unit buy or sell orders randomly, without any optimization or strategic reasoning.
\item \textbf{Fundamental-chartist-noise (FCN-) Agent}~\citep{fcn2}: FCN-Agent is a widely adopted agent model for LOB market simulations. It is primarily a mixture of fundamental and chartist investment strategies, incorporating heterogeneity in time horizons and risk aversion.
\item \textbf{adaptive FCN-Agent (adFCN-Agent)}~\citep{adaptive_fcn_agent}: adFCN-Agent is a variant of FCN-Agent, which adaptively selects between fundamental and chartist strategies based on their recent predictive accuracy.
\item \textbf{Ours (fixed)}: Our method without trait heterogeneity. All agents' trait factors are fixed with population mean.%as: $\forall j~~\alpha^j=2.0,\sigma^j=0.02,\gamma^j=0.95$.
\item \hashimoto{\textbf{Ours (uniform $\alpha^j$)}: Our method with uniformly distributed risk aversion term $\alpha^j$ over the same support of Ours. }
\end{itemize}
The FCN-Agent models only heterogeneity in preferences, whereas the adFCN-Agent with fixed risk aversion and time horizons across agents (adFCN-Agent (fixed)) models only learning. The key parameters of these ABMs were calibrated using the OT distance $\bar{OT}$, in the same manner as in our model. \hashimoto{The implementation details of these baselines are described in Appendix~\ref{App:baselines}.}

\subsection{Evaluation Metrics}
In addition to the OT distances, we assessed whether the synthetic data exhibited key stylized facts observed in real financial markets. Following \citet{stylized_facts} and \citet{realism}, we considered the following stylized facts:

\begin{itemize}
\item \textbf{Kurtosis}: The kurtosis of the stock return distribution is positive, indicating fat tails.
\item \textbf{Tail coefficient (Tail coef.)}: The Hill tail exponent~\citep{hill} is approximately three, reflecting the power-law behavior of extreme returns.
\item \textbf{Autocorrelation coefficient (Acorr coef.)}: The estimated coefficient $\hat{\zeta}_r^{\mathrm{acorr}}$ in the linear regression of $\log \mathrm{Corr}(|r_{t,i}|, |r_{t+\tau,i}|)$ on $\log \tau$ lies between 0 and 1, indicating the long memory property of absolute return series. Here, $\mathrm{Corr}(\cdot,\cdot)$ denotes the correlation operator.
\item \textbf{Volume-Volatility correlation (VV corr.)}: The correlation between the executed trading volume %within a one-minute interval
and the absolute return is positive. %over the same interval is positive.
\end{itemize}

%\hashimoto{The detailed calculation of these metrics are provided in Appendix~\ref{App:metrics}.}

\section{Results and Discussions}

This section presents and discusses our experimental results. \hashimoto{RQ1 analyzes how interacting agents with heterogeneous preferences and learning mechanisms organize into a market ecology characterized by differentiated trading roles, while RQ2 examines how the resulting market ecology gives rise to characteristic price dynamics observed in financial markets.}

\subsection{Behavioral Differentiations \hashimoto{and Self-Organized Market Ecology}}% Driven by Heterogeneous Preference}

\subsubsection{Behavioral Analyses}

\begin{figure*}[tbp]
  \centering
  % ---- 1st row (4 figs) ----
  \begin{subfigure}[b]{0.48\textwidth}
    \includegraphics[width=\linewidth]{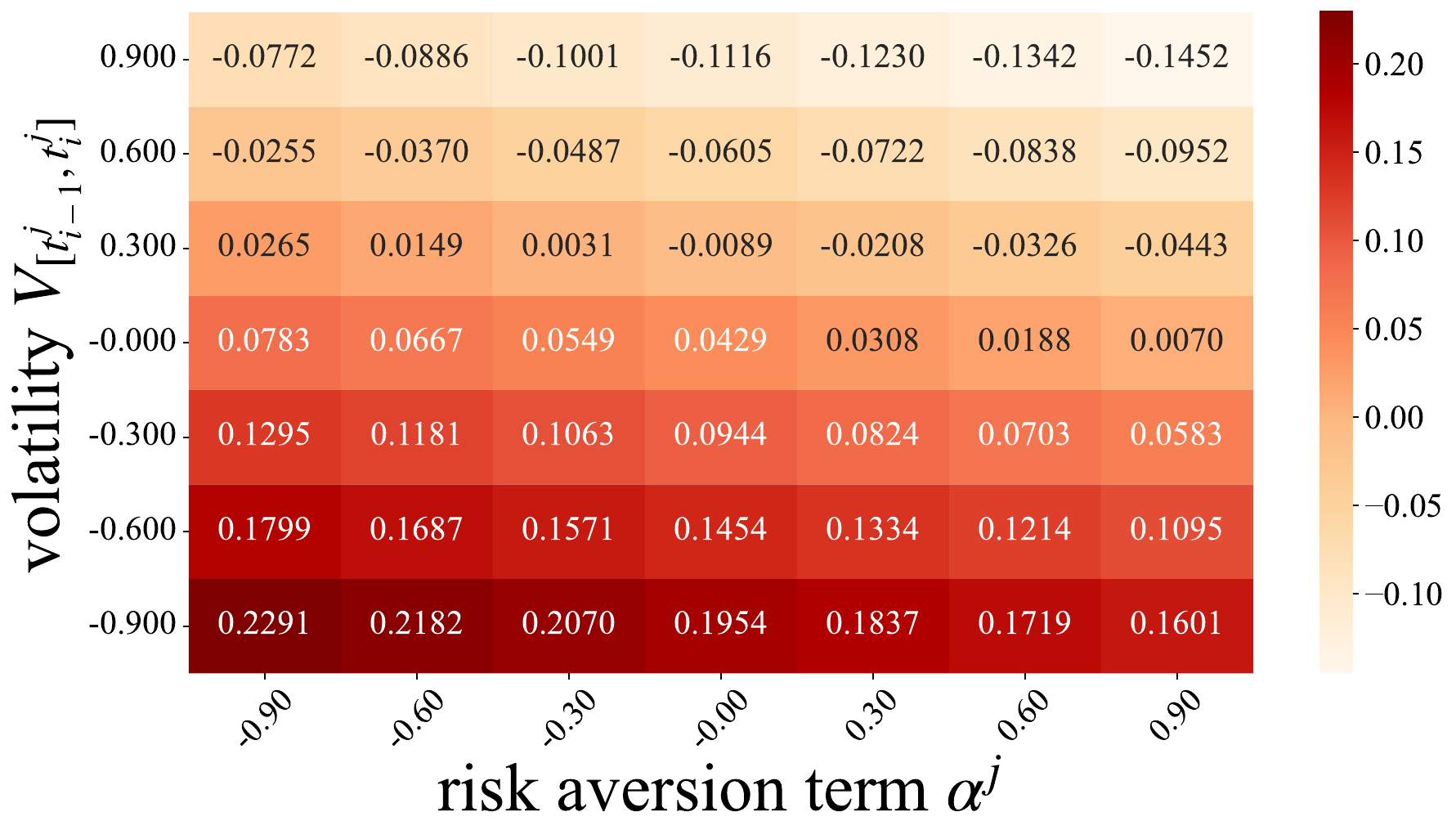}
    \caption{Ours}
    \label{Fig:normal}
  \end{subfigure}
  \begin{subfigure}[b]{0.48\textwidth}
    \includegraphics[width=\linewidth]{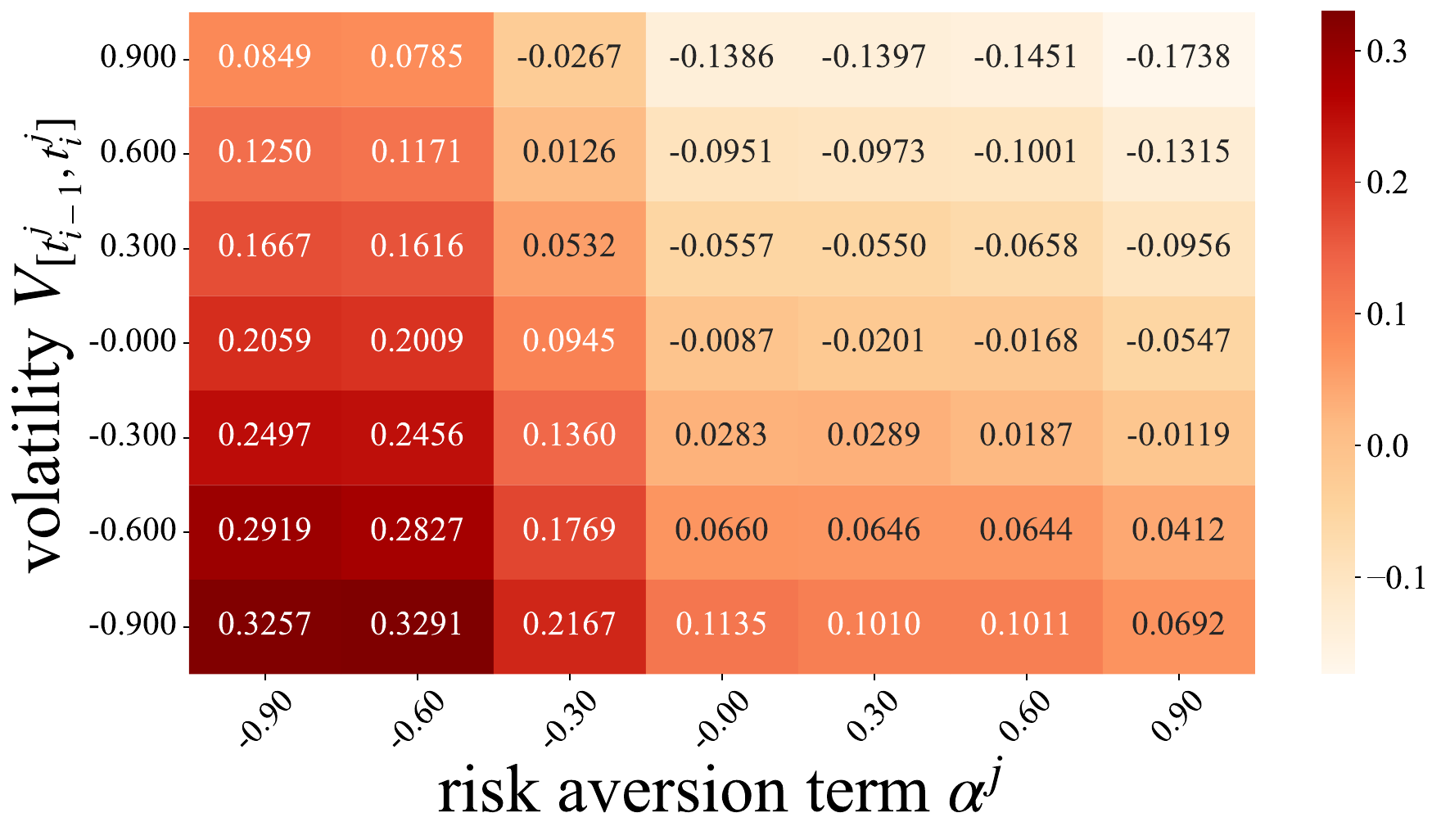}
    \caption{\hashimoto{Ours (uniform $\alpha^j$)}}
    \label{Fig:uniform}
  \end{subfigure}
  \caption{Heatmap of the scaled order volume $\tilde{v}_t^j$ derived from the obtained policy, with varying rescaled volatility $V_{[t_{i-1}^j, t_i^j]}$ and risk aversion term $\alpha^j$.}
  \label{Fig:heatmap_order_volume_given_alpha_volatility}
\end{figure*}

To investigate RQ1, we analyzed the learned policy obtained through our method to investigate how each trait factor contributes to shaping agent behavioral patterns. Figure~\ref{Fig:heatmap_order_volume_given_alpha_volatility} compares heatmaps illustrating the values of $\tilde{v}_t^j$ while varying the two observation components, $V_{[t_{i-1}^j, t_i^j]}$ and $\alpha^j$, between Ours and Ours (uniform $\alpha^j$). The values of the remaining variables were fixed. \hashimoto{Although Figures~\ref{Fig:heatmap_order_volume_given_alpha_volatility}(\subref{Fig:normal},\subref{Fig:uniform}) indicate similar tendency that agents tended to sell stocks more actively as their risk aversion increased and market volatility rose, they differ markedly in the structure of behavioral differentiation. In Figure~\ref{Fig:heatmap_order_volume_given_alpha_volatility}(\subref{Fig:normal}), trading behavior varies smoothly with $\alpha^j$, whereas in the uniform population (Figure~\ref{Fig:heatmap_order_volume_given_alpha_volatility}(\subref{Fig:uniform})) a sharp sign reversal emerges around $\alpha^j \approx -0.3$, reflecting a clear polarization of trading responses. Under the uniform distribution, agents with extreme values of $\alpha^j$ constitute a relatively larger fraction of the population, and their actions exert persistent influence on price formation and, consequently, on the reward structure faced by other agents. As a result, even agents with intermediate risk preferences are driven toward distinctly risk-seeking or risk-averse behaviors, leading to a clear bimodal polarization in trading responses. These findings demonstrate that, even for agents sharing the same value parameter $\alpha^j$, the structure of behavioral differentiation depends critically on the composition of the population where they inhabit---that is, on the relative positioning of their preferences within the population-level distribution.}

\begin{figure*}[tbp]
  \centering
  \begin{subfigure}[t]{0.46\textwidth}
    \includegraphics[width=\linewidth]{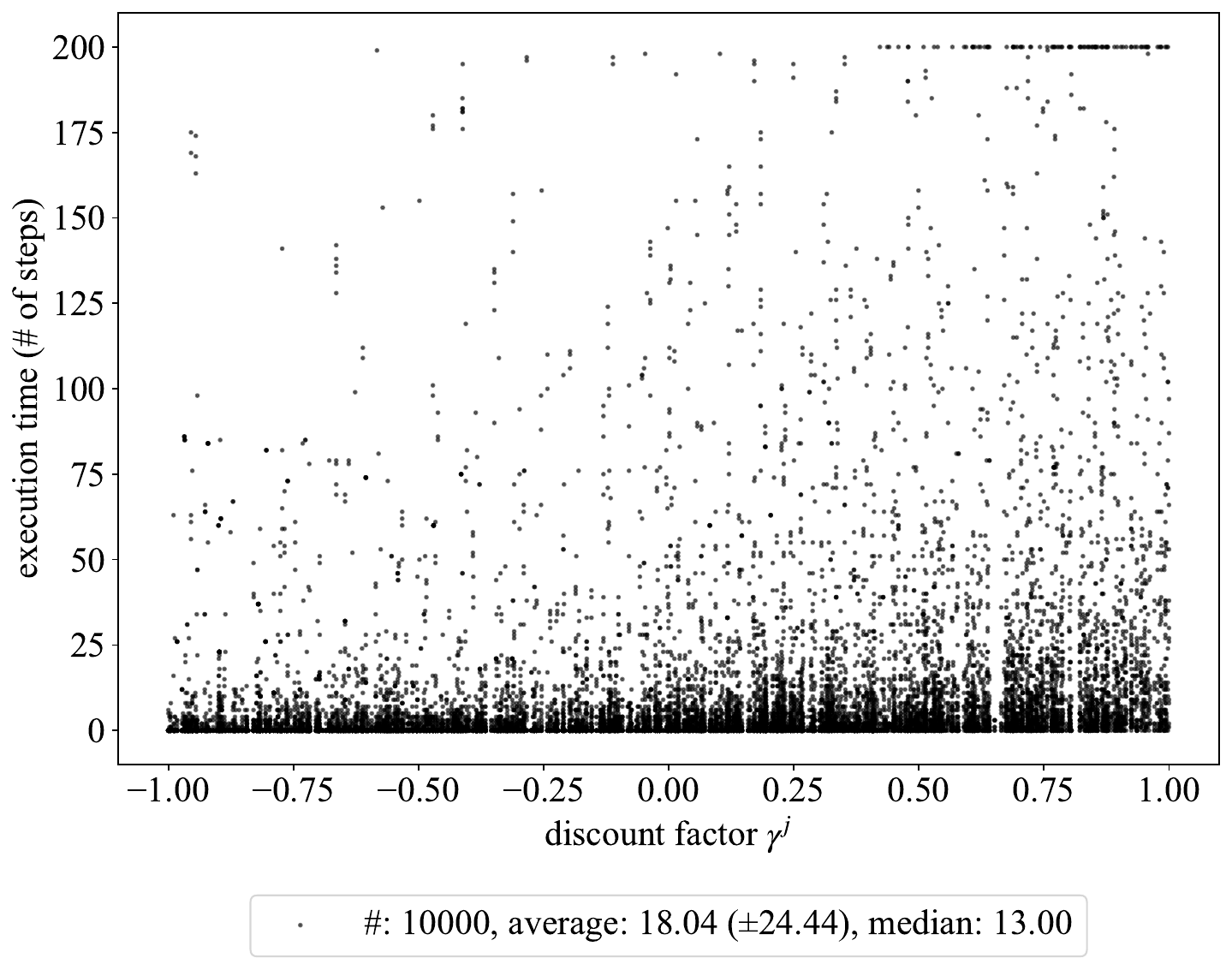}
    \caption{\hashimoto{Distribution of execution times for all submitted orders after the burn-in period. Summary statistics, including the sample size, mean, standard deviation, and median execution time, are reported in the figure legend.}}
    \label{Fig:scatter_execution_times}
  \end{subfigure}
  \begin{subfigure}[t]{0.50\textwidth}
    \includegraphics[width=\linewidth]{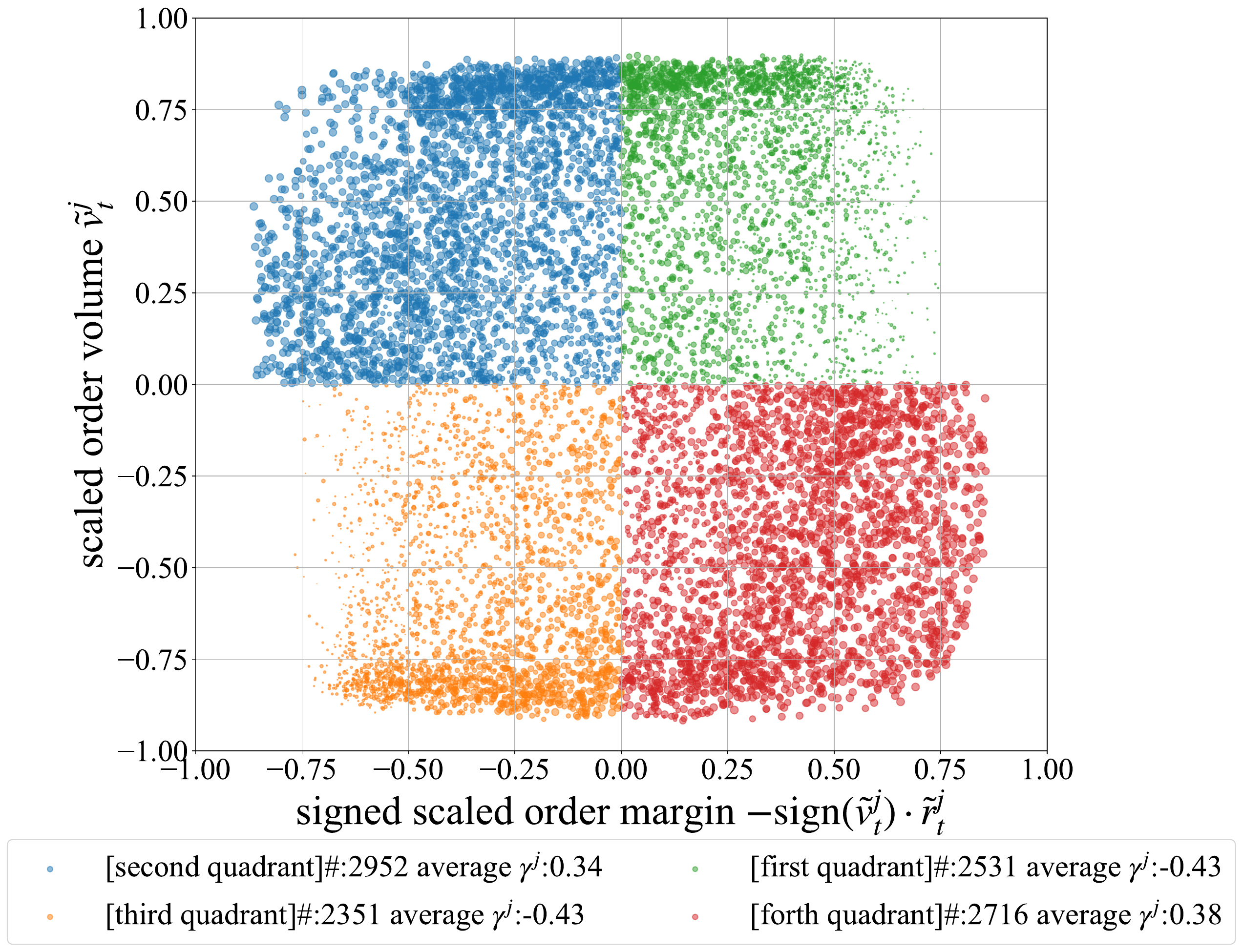}
    \caption{Scatter plot of agent action vectors from five simulations, with color coding for each quadrant. The sample size and mean discount factor $\gamma^j$ of each quadrant are reported in the figure legend. The size of each point represents the agent's $\gamma^j$.}
    \label{Fig:scatter_quadrants}
  \end{subfigure}
  \caption{\hashimoto{Scatter plots indicating the relationship between agents' discount factors and their trading actions.}}
  \label{Fig:scatter_actions}
\end{figure*}

\hashimoto{Figure~\ref{Fig:scatter_actions}(\subref{Fig:scatter_execution_times}) shows the distribution of execution times for all submitted orders after the burn-in period, with a finite time-in-force (TIF) of $200$ steps. Each point represents an individual order, plotted against the discount factor of the submitting agent. Half of the orders are executed within 13 steps, while a small fraction ($2.21\%$) remain unexecuted and expire after reaching the TIF limit. In addition, orders submitted by agents with higher discount factors $\gamma^j$ tend to persist longer in the LOB, indicating a greater propensity for resting orders among more long-horizon agents.} Figure~\ref{Fig:scatter_actions}(\subref{Fig:scatter_execution_times}) presents the scatter plot of agent actions $\tilde{v}_t^j$ and $\tilde{r}_t^j$ across five simulation trials. Agent actions are distributed across all quadrants of the action space, defined by order volume and margin. The second and fourth quadrants represent behavior consistent with price-sensitive trading---selling above the mid price or buying below it---while the first and third quadrants reflect the opposite: trading at disadvantageous prices. The average discount factors $\gamma^j$ differ systematically across these regions. Agents acting in the second and fourth quadrants tend to have higher $\gamma^j$ values, suggesting a more patient, forward-looking attitude. Conversely, agents in the first and third quadrants exhibit lower average discount factors, indicating a more myopic preference structure. This pattern suggests that myopic agents are more willing to accept unfavorable prices to execute their trades immediately, prioritizing execution speed over price optimization. \hashimoto{Importantly, the discount factor $\gamma^j$ does not directly prescribe agents’ action rules, but only determines how future rewards are weighted in their objective functions.
Nevertheless, through learning and repeated interaction in the endogenous market environment, agents with lower $\gamma^j$ systematically adopt strategies that favor immediate execution at unfavorable prices, effectively acting as liquidity consumers, while agents with higher $\gamma^j$ tend to submit more price-sensitive orders, functioning as liquidity providers. This role differentiation is not imposed ex ante, but emerges endogenously from the interaction between heterogeneous time preferences, learning dynamics, and market feedback. As agents adapt to the evolving behavior of others, differences in temporal valuation are translated into complementary trading roles, giving rise to a self-organized market ecology characterized by the coexistence of impatient liquidity demanders and patient liquidity providers.}

\begin{figure}[tbp]
  \centering
  \includegraphics[width=0.6\linewidth]{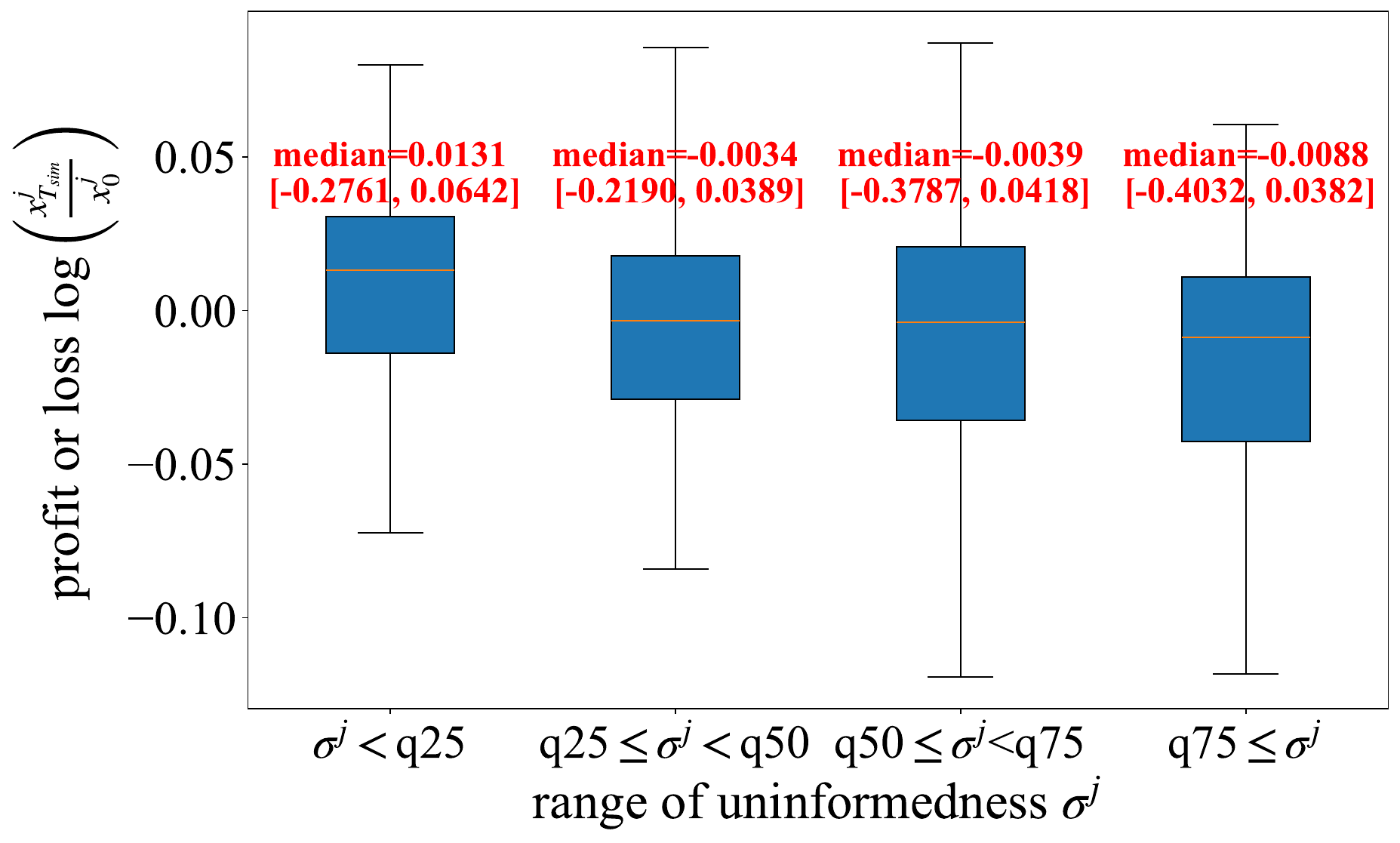}
  \caption{Box plot of agents’ trading performance over five runs. Agents are grouped by uninformedness $\sigma^j$ quartiles (q{quartile(\%)}) on the horizontal axis, and vertical axis shows log returns. %$\log x_{T_{sim}}^j - \log x_0^j$.
  Boxes indicate medians and 90\% ranges, showing the distribution of each trading performance.}
  \label{Fig:boxplot_uninformedness}
\end{figure}

Figure~\ref{Fig:boxplot_uninformedness} presents a box plot illustrating the relationship between agent uninformedness $\sigma^j$ and trading performance. Agents are grouped into quartiles based on their level of uninformedness, with lower $\sigma^j$ indicating better access to information. The plot reveals a negative trend: as uninformedness increases, the profit distribution shifts downward. This pattern demonstrates how differences in access to information influence agent behavior and ultimately lead to disparities in trading performance.

\subsubsection{Probing of the Shared-Policy}

\begin{figure*}[tbp]
  \centering
  \centering
  \includegraphics[width=\linewidth]{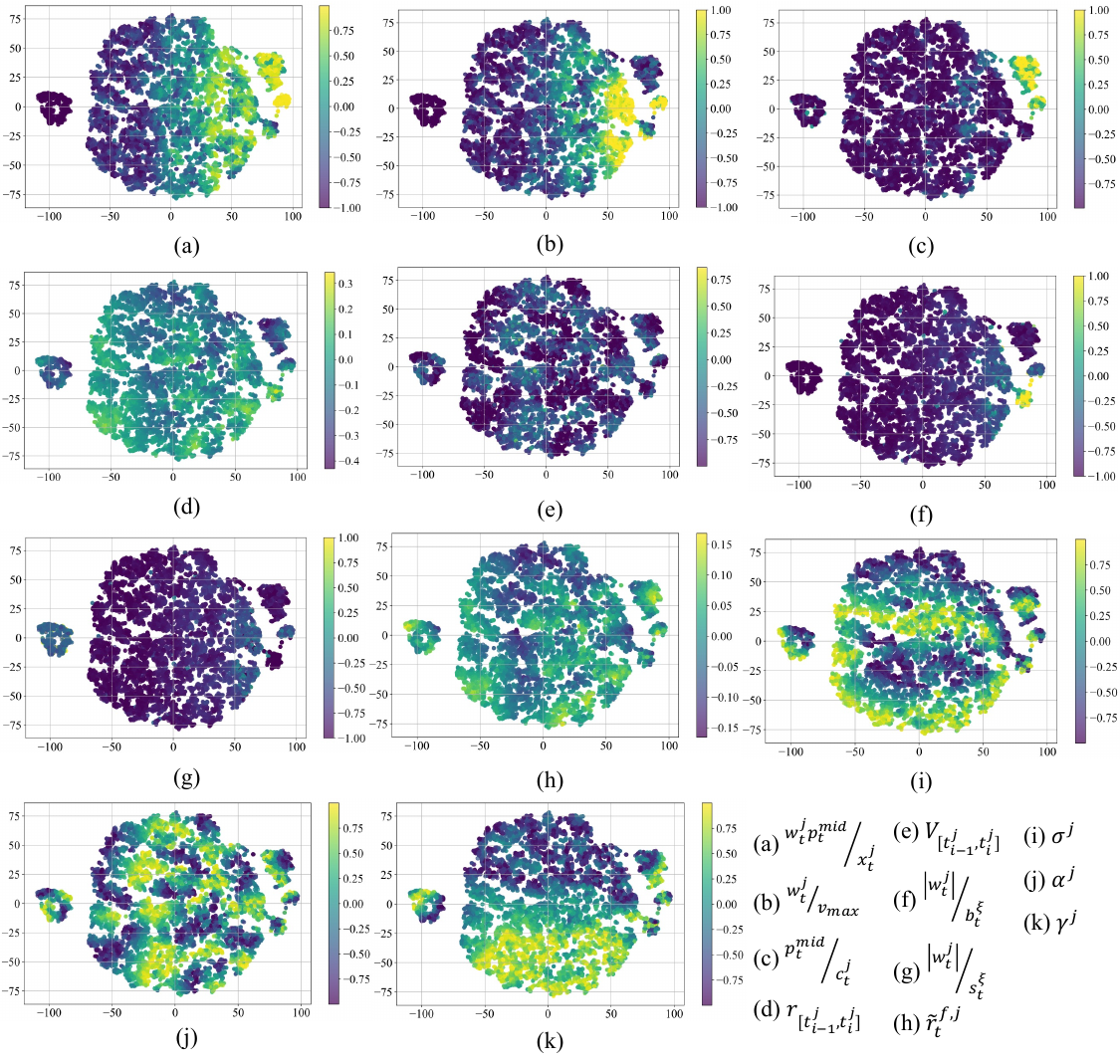}
  \caption{\hashimoto{t-SNE visualizations of internal representations in the second hidden layer, colored by various observation components.}}
  \label{Fig:tsne_observation_components}
\end{figure*}

To further support our claim to answer RQ1 that agent behaviors are differentiated according to their individual traits, we conducted a probing analysis on the learned shared-policy network. Figures~\ref{Fig:tsne_observation_components} show the t-SNE visualizations of the internal representations extracted from the second hidden layer of the shared-policy network. These representations were collected across five independent evaluation episodes. These visualizations illustrate how different observation components are embedded in the learned internal representations. As shown in Figures~\ref{Fig:tsne_observation_components}~(i,j,k), the internal representations are systematically organized according to agent-specific traits. In the three visualizations, the data points are distinctly clustered by their corresponding trait values. This consistent trait-wise separation in the second hidden layer suggests that the network implicitly encodes a latent notion of {\em who the agent is} based on its individual preferences.

Figures~\ref{Fig:tsne_observation_components}~(i,j,k) show smooth, continuous gradients along the latent dimensions when colored by each agent’s trait parameters, indicating that the shared-policy develops an internal representation that smoothly encodes who the agent is---i.e., its identity defined by long-lasting traits. In contrast, Figures~\ref{Fig:tsne_observation_components}~(c,f,g) reveal more clustered and discrete structures. These features correspond to the agent’s transient state, such as its current wealth or market liquidity. Notably, the reward function includes components like the cash shortage penalty and illiquidity penalty, which are implemented as piecewise constant functions---yielding sharply negative rewards when certain thresholds are crossed. This structural design likely induces discrete transitions in the agent’s internal representation of situational context. Together, these results suggest that the learned shared-policy not only encodes each agent’s stable identity (trait-driven preferences) in a smooth manner, but also dynamically reacts to short-term agent or market states in a thresholded, categorical fashion. In doing so, the policy integrates both long-term traits and short-term states to adaptively adjust agent behavior, supporting the emergence of diverse and context-aware strategies.

\subsubsection{Ablation Study}

\hashimoto{We conducted an ablation study to examine whether preference heterogeneity serves as a fundamental source of interdependent coexistence, which lies at the core of ecosystem dynamics.} Specifically, we ran five independent simulations for each setting and computed the sum of discounted cumulative utilities across all agents, denoted as $\mathcal{U}$:
\begin{align}
\mathcal{U}=\sum_{j=1}^n\sum_{i=1}^{\iota_j}(\gamma^j)^iu_{t_i^j}^j
\end{align}
Two types of ablations were implemented: (i) homo, where the focal trait of agents is fixed to its population mean, thereby eliminating heterogeneity, and (ii) masked, where the corresponding trait input to the shared-policy network is masked. Each ablated model is trained entirely from scratch. 
As reported in Table~\ref{Tab:ablation_heterogeneity}, removing heterogeneity in key preference parameters such as risk aversion ($\alpha^j$) and time discounting ($\gamma^j$) leads to a substantial reduction in overall social utility. This result highlights that when all agents share the same risk attitude or time horizon, the system loses the diversity of demand–supply matching that is otherwise enabled by heterogeneous preferences. For example, risk-tolerant agents and risk-averse agents, or short-term oriented and long-term oriented agents, complement each other in ways that increase aggregate utility.
The fact that utility decreases in both homo and masked settings provides evidence that preference-driven behavioral differentiation plays a critical role in enhancing population utility. Each agent adapts its strategy in alignment with its individual preference, leading to a diversified financial ecosystem in which agents specialize in distinct niches. Preference heterogeneity, when coupled with learning, induces functional complementarity~\citep{complementarity} among such agent niches---an organismic order in which no single subpopulation can reproduce the system’s performance alone.

\begin{table}[tbp]
\centering
\caption{Ablation study on preference heterogeneity. We report the average of the discounted cumulative utility $\mathcal{U}$ (mean $\pm$ std. over five trials). }
\begin{tabular}{cc}
\toprule
Model & Avg. Cum. Disc. Utility\\
\midrule
Ours & $1016.88~(\pm41.22)$ \\
Ours (homo-$\alpha^j$) & $846.58~(\pm29.85)$ \\
Ours ($\alpha^j$ masked) & $799.32~(\pm83.47)$ \\
Ours (homo-$\gamma^j$) & $861.98~(\pm51.02)$ \\
Ours ($\gamma^j$ masked) & $606.58~(\pm102.39)$ \\
Ours (homo-\hashimoto{$\sigma^j$}) & $1009.98~(\pm44.00)$ \\
Ours (\hashimoto{$\sigma^j$} masked) & $916.53~(\pm20.87)$ \\
\bottomrule
\end{tabular}
\label{Tab:ablation_heterogeneity}
\end{table}

\subsection{Emergence of Market Dynamics}

\begin{table*}[tbp]
\centering
\caption{Comparison of stylized facts and OT distances from real data across financial ABMs. A check mark ($\checkmark$) denotes conformity to each stylized fact. OT columns show mean distances between synthetic and real point clouds (standard deviations in parentheses); the best values are in \textbf{bold}.}
\label{Tab:benchmark_results}
%
% --------------------- Subtable A: Stylized Facts ---------------------
\begin{subtable}{\textwidth}
\centering
\caption*{(a) Stylized Facts}
\begin{tabular*}{\textwidth}{@{\extracolsep{\fill}}lcccc}
\toprule
          & Kurtosis & Tail coef. & Acorr coef. & VV corr. \\
\midrule
Real data & 7.79$\checkmark$ & 2.96$\checkmark$ & 0.71$\checkmark$ & 0.43$\checkmark$ \\
\midrule
ZI-Agent & 1.19$\checkmark$ & 3.79 & \textemdash{} & 0.04$\checkmark$ \\
FCN-Agent & 6.25$\checkmark$ & 3.23 & 1.11 & 0.08$\checkmark$ \\
adFCN-Agent (fixed) & 31.38$\checkmark$ & 2.56 & 1.01 & 0.01$\checkmark$ \\
adFCN-Agent & 10.38$\checkmark$ & 2.82$\checkmark$ & 0.77$\checkmark$ & 0.05$\checkmark$ \\
\midrule
Ours & 8.65$\checkmark$ & 2.99$\checkmark$ & 0.92$\checkmark$ & 0.16$\checkmark$ \\
Ours (fixed) & 12.89$\checkmark$ & 3.44 & 1.05 & 0.16$\checkmark$ \\
\bottomrule
\end{tabular*}
\end{subtable}

\vspace{0.8em}

% --------------------- Subtable B: OT Distances ---------------------
\begin{subtable}{\textwidth}
\centering
\caption*{(b) OT Distances from Real Data}
\begin{tabular*}{\textwidth}{@{\extracolsep{\fill}}lccc}
\toprule
          & $OT^r~(\times10^{-1})$ & $OT^t~(\times10^{-2})$ & $OT^{as}$ \\
\midrule
Real data & \textemdash{} & \textemdash{} & \textemdash{}\\
\midrule
ZI-Agent & $1.80~(\pm0.79)$ & $1.00~(\pm0.52)$ & $0.34~(\pm0.03)$\\
FCN-Agent & $1.34~(\pm0.31)$ & $0.91~(\pm0.61)$ & $0.30~(\pm0.02)$\\
adFCN-Agent (fixed) & $2.34~(\pm0.45)$ & $1.05~(\pm0.62)$ & $0.26~(\pm0.02)$\\
adFCN-Agent & $0.65~(\pm0.27)$ & $0.48~(\pm0.30)$ & $0.26~(\pm0.04)$\\
\midrule
Ours & $\textbf{0.37}~(\pm0.32)$ & $\textbf{0.32}~(\pm0.24)$ & $\textbf{0.10}~(\pm0.03)$ \\
Ours (fixed) & $1.40~(\pm0.89)$ & $0.65~(\pm0.56)$ & $0.19~(\pm0.02)$ \\
\bottomrule
\end{tabular*}
\end{subtable}

\end{table*}
To address RQ2, Table~\ref{Tab:benchmark_results} compares our method with baseline ABMs.
FCN-Agent incorporates only heterogeneous preferences, while adFCN-Agent (fixed) and Ours (fixed) model only the learning mechanism. In contrast, adFCN-Agent and Ours implement both heterogeneity and learning simultaneously. The results show that our model and adFCN-Agent successfully reproduce all the key stylized facts, whereas models that adopt only one of the two mechanisms fail to capture certain empirical regularities. This finding suggests that both heterogeneous preferences and adaptive learning are indispensable drivers of collective dynamics in financial markets.
Moreover, the OT distances reveal a clear advantage of our method. Across all three metrics, Ours achieves the lowest distance from real data, outperforming not only the purely heterogeneity-based and purely learning-based baselines, but also the adFCN-Agent. \hashimoto{To summarize, behavioral differentiation arising from heterogeneous preferences and interactive learning gives rise to realistic financial market dynamics through agent interactions. By jointly modeling preference heterogeneity and learning mechanisms, our framework enables the self-organization of a market ecology in which differentiated trading roles emerge endogenously. This self-organized structure provides a foundation for reproducing characteristic features of real financial markets, such as fat-tailed returns and volatility clustering. In this sense, the improved performance of our model relative to adFCN-Agent stems not from a single modeling component, but from an integrated representation of adaptation and interaction across individual and collective scales.}

\begin{figure}[tbp]
  \centering
  \includegraphics[width=0.6\linewidth]{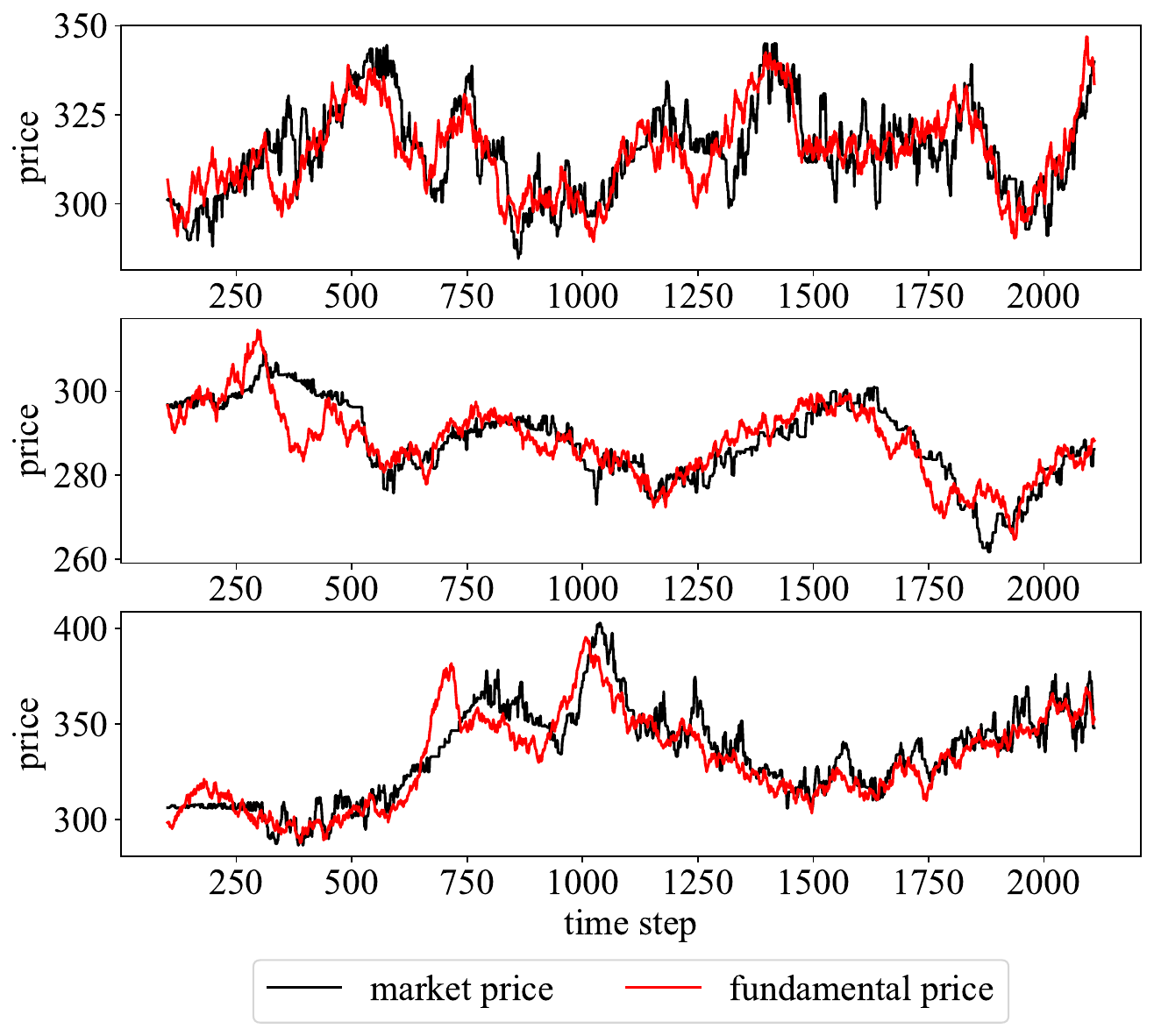}
  \caption{Representative examples of simulated market price series generated by our method.}
  \label{Fig:simulated_prices}
\end{figure}

Figure~\ref{Fig:simulated_prices} shows examples of simulated market prices generated by our method. Market prices generally follow fundamental values, indicating that agents have collectively learned the norm that prices should reflect fundamentals. At the same time, short-term deviations and fluctuations suggest market noise and diverse agent behaviors. Overall, the model strikes a balance between agent-level diversity and coherent market structure. Agents act differently, yet their interactions produce organized price dynamics.

\section{Conclusion}
\hashimoto{This study approached financial markets as an artificial life-like system, using them as a constructive testbed to explore how adaptive collective organization emerges from the interaction of heterogeneous preferences and learning mechanisms.} In our MARL-based ABM, agent preferences are embedded in both observations and rewards, and agents learn a shared-policy that yields behaviors aligned with their preferences.
Experiments showed that (i) agents differentiated their behavior according to preferences \hashimoto{through interactive learning}, leading to niche specialization, and (ii) interactions among differentiated agents reproduced realistic market dynamics. \hashimoto{These findings highlight the importance of the integration of heterogeneous preferences and adaptive learning for shaping market ecology as a foundation of the emergence of realistic price dynamics, providing a computational account of how adaptive market behavior can arise from interacting, learning agents.}

% hierarchical emergence -> Loop of Emergence
% Future research may extend such a hierarchical emergence to the {\em loop of emergence} by incorporating micro-macro loops in which agents repeatedly adapt to and shape evolving environments. Such a co-evolutionary dynamic enables the simulation as nonequilibrium systems, capturing more realistic patterns of structural change in financial markets.

\section{Limitations}
\paragraph{\textbf{\hashimoto{Limited Scope of Trait Distributions and Differentiation Patterns}}}
\hashimoto{In this study, agent traits such as risk aversion, discounting, and uninformedness are drawn from relatively simple distributions, primarily focusing on unimodal or uniform cases. While this design facilitates controlled analysis of how trait heterogeneity interacts with learning and market interaction, it does not capture the full diversity of trader distributions reported in behavioral economics studies, where preference distributions can be skewed, heavy-tailed~\citep{distribution_risk_averion}, related to income~\citep{income_risk_aversion}, or even time-varying~\citep{time_varying_risk_aversion}. Extending the framework to incorporate multi-modal trait distributions or correlated preferences would provide a richer testbed for exploring how more complex populations influence the self-organization of market ecologies.}

\paragraph{\textbf{\hashimoto{Reliance on a Shared-Policy}}} \hashimoto{
Using a shared-policy~\citep{shared_policy_learning,learning_and_heterogeneity3} involves a trade-off compared to learning individual policies for each agent. A shared-policy reduces learning-related control variables by training all agents under the same function approximator and optimization process, which is particularly useful in our setting for examining how preference heterogeneity is translated into differentiated behavior while holding the learning mechanism constant. At the same time, shared-policy learning may limit exploration and introduce interference effects due to pooled experiences from heterogeneous agents. Individual policies could allow more independent adaptation and a wider range of behaviors, but they also introduce additional variability from factors such as initialization and stochastic learning dynamics, making it harder to isolate the role of preference heterogeneity, which is the central focus of this study. For these reasons, we adopt a shared-policy framework as a deliberate modeling choice rather than a claim of optimality, and leave exploration under individual policies as future work.
}

\paragraph{\textbf{\hashimoto{Lack of Selection and Turnover in Market Ecologies}}}
\hashimoto{While our results demonstrate that heterogeneous preferences and adaptive learning are sufficient to construct a coherent market ecology and reproduce realistic price dynamics, they do not yet capture how market ecologies themselves evolve or transform over time.
Addressing this limitation requires extending the framework to include mechanisms of selection and replacement, whereby agents adapted to a particular environment may be eliminated or disadvantaged as market conditions change, allowing new behaviors and structures to emerge.
Incorporating such evolutionary dynamics is an important direction for future work toward modeling continuously evolving financial market ecologies.}

\section*{Acknowledgments}
This work was supported by JSPS KAKENHI Grant Number JP25KJ1124.

\bibliographystyle{plainnat}
\bibliography{NAME}

\appendix

\section{\hashimoto{Our Method: POMDP Setup}}\label{App:pomdp}

\begin{table}[htbp]
\centering
\caption{Simulation settings and hyperparameters used in the proposed framework.}
\label{Tab:simulation_settings}
\begin{tabular}{lll}
\toprule
\textbf{Parameter} & \textbf{Notation} & \textbf{Value} \\
\midrule
\# of agents     & $n$       & 200\\
Time to order expiration & \textemdash{} & 200\\
\# of time steps & $T_{sim}$ & 2{,}110\\
Expected initial asset volume & $w$ & 20\\
Expected initial cash amount  & $c$ & 15{,}000.00\\
\makecell[l]{Maximum LOB depth\\
~~for existing order \\
~~volume calculation}      & $\xi$ & 0.05\\
\makecell[l]{Decay rates for existing \\
~~buy/sell order volume}           & $\omega^b,\omega^s$      & 100\\
Maximum order volume          & $v_{max}$ & 20\\
\makecell[l]{Maximum scaled order\\
~~price margin}                     & $r_{max}$  & 0.05\\   
Expected uninformedness       & $\mu^\sigma$   & 0.02\\
Expected risk aversion term   & $\mu^\alpha$   & 2.00\\
Maximum time discount factor  & $\gamma^{max}$ & $9.99\times10^{-1}$\\
\makecell[l]{Penalty weight\\
~~for short selling} & $\beta^{short}$ & 0.10\\
\makecell[l]{Penalty weight\\
~~for cash shortage} & $\beta^{cash}$  & 0.10\\
Penalty weight for illiquidity   & $\beta^{illiquidity}$ & $0.05\times10^{-1}$\\
\makecell[l]{Penalty weight for\\
~~deviation from fundamentals}   & $\beta^{fundamentel}$ & 0.20\\
Utility scaling factor        & $\omega^u$ & $0.06\times10^{-3}$\\
\makecell[l]{Scaling factor for\\
~~order volume imbalance}     & $\omega^l$ & 1.00\\
Initial fundamental price     & $p_0^f$    & 300.00\\
\bottomrule
\end{tabular}
\end{table}

Table~\ref{Tab:simulation_settings} summarizes the experimental setup of the proposed multi-agent simulation framework. In each simulation, for the first $100$ steps and from step $1,100$ to $1,110$, only order placement is allowed; no order execution occurs. This corresponds to the pre-opening and midday lunch break periods in the Tokyo Stock Exchange, during which orders can be submitted but matching is temporarily suspended. The fundamental price process $p_t^f$ follows a geometric Brownian motion (GBM) with a fixed drift of $0.00$ and volatility sampled uniformly from the range $[0.00,0.03\times10^{-1}]$. All observation vectors were min-max normalized to lie within the interval $[-1,1]$.

To assess the realism of the simulated markets, we conducted 300 independent simulation runs and converted the resulting synthetic data into the same structured format as the real data. Specifically, we applied the time deformation technique proposed by \citet{ot_based_simulation_evaluation} to transform the simulation outputs into one-minute bar series of prices and volumes.

\section{\hashimoto{Our Method: Training Details}}\label{App:training}

\begin{table}[htbp]
\centering
\caption{Hyperparameter details of training and calibrating proposed algorithm in the experiment.}
\label{Tab:training_details}
\begin{tabular}{lll}
\toprule
\textbf{Parameter} & \textbf{Notation} & \textbf{Value} \\
\midrule
Rollout length     & $T_{rollout}$      & 1024\\
Batch size         & \textemdash{}      & 512\\
\# of updates per rollout & \textemdash{} & 5\\
\makecell[l]{Learning rates for the\\
~~actor/critic networks}   & $\beta^{actor},\beta^{critic}$ & $0.01\times10^{-2}$\\
\makecell[l]{Clipping parameter for\\
~~the surrogate objective} & $\epsilon$ & 0.80\\
Maximum gradient norm & \textemdash{}   & 0.50\\
\makecell[l]{Log-standard deviation\\
~~of squashed Gaussian policy} & \textemdash{} & -1.00\\
OT weight for return & $\omega^r$ & 1.00\\
OT weight for tail return & $\omega^t$ & 2.00\\
\makecell[l]{OT weight for\\
~~absolute return series} & $\omega^{as}$ & 4.00\\
\bottomrule
\end{tabular}
\end{table}

In our experiment, the shared-policy was trained using proximal policy optimization~\citep{ppo}. Both the actor and critic networks ($\pi_{\bm{\theta}}$ and $C_{\bm{\phi}}$) were implemented as three-layer neural networks with $512$ hidden units per layer and hyperbolic tangent activation functions. The actor network output followed a squashed Gaussian distribution~\citep{sac}. Consistent with best practices reported by \citet{ppo_empirical_study}, both networks were orthogonally initialized, and the final layer of the actor was initialized with weights scaled down by a factor of 100. During training, rewards stored in each agent’s rollout buffer $D_j$ were normalized by their sample standard deviation and clipped to the range $[-5,5]$. Additional hyperparameter settings are provided in Table~\ref{Tab:training_details}.

\begin{table}[htbp]
\centering
\caption{Hyperparameter candidates for calibration.}
\begin{tabular}{ll}
\midrule
Parameter          & Candidate values\\
\midrule
$\lambda^{\sigma}~(\times10^{-1})$ & $0.00$, $0.02$, $0.04$, $0.06$, $0.08$, $0.10$\\
$\lambda^{\alpha}$ & $0.00$, $0.20$, $0.40$, $0.60$, $0.80$, $1.00$\\
$\lambda^{\gamma}$ & $0.75$, $0.80$, $0.85$, $0.90$, $0.95$, $1.00$\\
\midrule
\end{tabular}
\label{Tab:calibrated_params}
\end{table}

In the experiment, we searched the hyperparameters $\lambda^\sigma$, $\lambda^\alpha$, and $\lambda^\gamma$ using OT-based calibration method, as described in the main paper. Table~\ref{Tab:calibrated_params} shows the candidate values to be searched. As a result of the calibration, the selected hyperparameters were $\lambda^\sigma=0.06\times10^{-1}$, $\lambda^\alpha=0.80$, and $\lambda^\gamma=0.90$.

\section{\hashimoto{Baselines}}\label{App:baselines}

\subsection{Zero intelligence (ZI-) Agent}
In financial market simulations, ZI-Agents serve as a behaviorally naive benchmark. They submit one unit buy or sell orders randomly, without any optimization or strategic reasoning. The ZI-Agent $j$'s order price $p_t^j$ is randomly determined as:
\begin{align}
p_t^j=p_t^{mid}\exp(r_t^j),~~r_t^j\sim\mathcal{N}\left(0,(\sigma^n)^2\right)
\end{align}
where $\sigma^n\in\mathbb{R}_+$ is a standard deviation of ZI-Agents' order prices.

\subsection{Fundamental-chartist-noise (FCN-) Agent}
\citet{fcn} proposed FCN-Agent model. FCN-Agent $j$ submits an order specifying a signed order volume $v_t^j \in \mathbb{Z}$ and an order price $p_t^j \in \mathbb{R}_+$. The sign of $v_t^j$ indicates whether the agent intends to buy or sell. The FCN-Agent $j$ predicts future stock returns $\hat{r}_{t+\tau_t^j}^j$ and price $\hat{p}_{t+\tau_t^j}^j$ given the current time steps $t$, market price $p_t$, and fundamental price $p_t^f$ as:
{\small
\begin{align}
    \hat{r}_{t+\tau_t^j}^j &= \frac{1}{w_t^{j,f} + w_t^{j,c} + w_t^{j,n}} \left( \frac{w_t^{j,f}}{\tau^{j,f}} \log \frac{p_t^f}{p_t} 
    + \frac{w_t^{j,c}}{\tau_t^j} \log \frac{p_t}{p_{t-\tau_t^j}} 
    + w_t^{j,n} \epsilon_t \right) \label{Eq:fcn_return} \\
    \hat{p}_{t+\tau_t^j}^j &= p_t \exp\left( \tau_t^j \hat{r}_{t+\tau_t^j}^j \right) \label{Eq:fcn_price}
\end{align}}
Here, $\tau_t^j\in\mathbb{N}$ denotes the time window size of agent $j$ at time $t$. $w_t^{j,f},w_t^{j,c},w_t^{j,n}\in\mathbb{R}_+$ are the coefficients corresponding to the three components described below, randomly determined for each agent at the beginning of the simulation such that $w_1^{j,f}\sim Ex(\lambda^f),~~ w_1^{j,c}\sim Ex(\lambda^c),~~ w_1^{j,n}\sim Ex(\lambda^n)$. Here, $Ex(\lambda)$ indicates an exponential distribution with expected value $\lambda\in\mathbb{R}_+$. Unless otherwise specified, these three coefficients are time-invariant: $\forall t~~ w_t^{j,f}=w_1^{j,f},w_t^{j,c}=w_1^{j,c},w_t^{j,n}=w_1^{j,n}$. The first term in Equation~(\ref{Eq:fcn_return}) $\frac{w_t^{j,f}}{\tau^{j,f}} \log \frac{p_t^f}{p_t} $ is called the fundamental trader component, implying that agent $j$ predicts that market price $p_t$ will reverse to fundamental price $p_t^f$ in $\tau^{j,f}$ steps. $\tau^{j,f}\in\mathbb{N}$ denotes the mean-reversion time of the agent $j$. The second term $\frac{w_t^{j,c}}{\tau_t^j} \log \frac{p_t}{p_{t-\tau_t^j}} $ is called the chartist trader component, which reflects past price movements in return predictions. The last term $w_t^{j,n} \epsilon_t$ is called the noise trader component. $\epsilon_t$ is Gaussian noise with a mean $0$ and variance $(\sigma^n)^2$. FCN-Agent was named after its three components, fundamental, chartist, and noise traders, which together form its return prediction model.

\citet{fcn2} proposed the order decision of the FCN-Agent based on price prediction $\hat{p}_{t+\tau_t^j}^j$, holding cash amount $c_t^j\in\mathbb{R}$, holding stock position $w_t^j\in\mathbb{Z}$, and the risk-aversion term $\alpha_t^j\in\mathbb{R}_+$ by assuming the constant absolute risk aversion (CARA) utility function $\mathcal{U}_t^j$:
\begin{align}
\mathcal{U}_t^j=-\exp\left\{-\alpha_t^j(w_t^jp_t+c_t^j)\right\}\label{Eq:cara}
\end{align}
where $\forall j~~ c_1^j\sim U(c_{min},c_{max}),~w_1^j=\lceil w^j\rceil,~w^j\sim U(w_{min},w_{max}),$\\ $c_{min},c_{max},w_{min},w_{max}\in\mathbb{R}_+$. $\lceil\cdot\rceil$ denotes the ceiling function and $U(c_{min},c_{max})$ represents the uniform distribution from range $[c_{min},c_{max}]$. $\alpha_t^j$ and $\tau_t^j$ are set as:
{\small
\begin{align}
\forall t,j~~ \alpha_t^j=\alpha\frac{\alpha^{diff}+w_t^{j,f}}{\alpha^{diff}+w_t^{j,c}},~\tau_t^j=\left\lceil\tau\frac{\tau^{diff}+w_t^{j,f}}{\tau^{diff}+w_t^{j,c}}\right\rceil\label{Eq:alpha_tau}
\end{align}}
where $\alpha,\tau\in\mathbb{R}_+$ are the reference levels of $\alpha_t^j$ and $\tau_t^j$, and $\alpha^{diff},\tau^{diff}\in\mathbb{R}_+$ are external parameters that control the range of variation in $\alpha_t^j$ and $\tau_t^j$ for each agent.

The agent decides their $p_t^j$ and $v_t^j$ to maximize their expected future utility $\mathbb{E}_t[\mathcal{U}_{t+\tau_t^j}^j]$, where $\mathbb{E}_t[\cdot]$ stands for the expected mean conditional to the information available at time $t$.

\subsection{adaptive FCN-Agent}
The adFCN-Agent adaptively chooses between the two forecasting rules---fundamentalist and chartist---based on their past predictive performance, thereby learning sequentially which rule is more effective at a given time. The adFCN-Agent modifies its fundamental and chartist weights $w_t^{j,f}, w_t^{j,c}$ by the following procedure. At time step $t$, the return predictions of fundamental trader at the time of $t-\tau^{j,f}$, denoted as $\hat{r}_{t-\tau^{j,f}}^f$ and the realized return from $t-\tau^{j,f}$, denoted as $r_{[t-\tau^{j,f},t]}$ is calculated as follows.
\begin{eqnarray}
\hat{r}_{t-\tau^{j,f}}^f=\log\frac{p_{t-\tau^{j,f}}^f}{p_{t-\tau^{j,f}}},~r_{[t-\tau^{j,f},t]}=\log\frac{p_t}{p_{t-\tau^{j,f}}}\label{Eq:f_pred_obs_return}
\end{eqnarray}
Similarly, we define the predictive and realized return of chartist trader, denoted as $\hat{r}_{t-\tau_t^j}^c$ and $r_{[t-\tau_t^j,t]}$, respectively.
\begin{eqnarray}
\hat{r}_{t-\tau_t^j}^c=\log\frac{p_{t-\tau_t^j}}{p_{t-2\tau_t^j}},~r_{[t-\tau_t^j,t]}=\log\frac{p_t}{p_{t-\tau_t^j}}\label{Eq:c_pred_obs_return}
\end{eqnarray}
The weight update function $f_w:\mathbb{R}_+\times\mathbb{R}\times\mathbb{R}\rightarrow\mathbb{R}_+$ calculates the updated weight given current weight $w$, predictive return $\hat{r}$, and realized return $r$ as follows.
\begin{align}
f_w(w,\hat{r},r|\eta, w^{max})=\begin{cases}
w + \eta\cdot100\cdot|r|\cdot(w^{max}-w) & \mathrm{if}~~\mathrm{sign}(\hat{r}r)=1\\
w-\eta\cdot100\cdot(|r|+|\hat{r}|)\cdot w & \mathrm{otherwise}
\end{cases}\label{Eq:f_w}
\end{align}
Here, $\eta\in\mathbb{R}_+$ is learning rate. $w^{max}$ indicates the maximum value of $w$. Equation~(\ref{Eq:f_w}) represents that $f_w$ increases the weight if the sign of predictive return $\hat{r}$ matches that of realized return $r$ and decreases it otherwise. Using $\tilde{w}_t^{j,f}=f_w(w_t^{j,f},\hat{r}_{t-\tau^{j,f}}^f,r_{[t-\tau^{j,f},t]}|\eta)$ and $\tilde{w}_t^{j,c}=f_w(w_t^{j,c},\hat{r}_{t-\tau_t^{j,c}},r_{[t-\tau_t^j,t]}|\eta)$, the adFCN-Agent updates their weights as:
\begin{eqnarray}
w_t^{j,f}&\leftarrow&(w_t^{j,f}+w_t^{j,c})\frac{\tilde{w}_t^{j,f}}{\tilde{w}_t^{j,f}+\tilde{w}_t^{j,c}}\\
w_t^{j,c}&\leftarrow&(w_t^{j,f}+w_t^{j,c})\frac{\tilde{w}_t^{j,c}}{\tilde{w}_t^{j,f}+\tilde{w}_t^{j,c}}
\end{eqnarray}
This update mechanism allows the adFCN-Agent to dynamically adjust $w_t^{j,f}$ and $w_t^{j,c}$ according to the latest accuracy of their predictive returns without changing noise trader ratio: $(w_t^{j,f}+w_t^{j,c}):w_t^n$. Let us define the chartist ratio $r_t^c$ in the entire market as:
\begin{eqnarray}
r_t^c=\mathrm{Median}\left(\frac{w_t^{j,c}}{w_t^{j,f}+w_t^{j,c}}\right)\label{Eq:r_c}
\end{eqnarray}
where $\mathrm{Median}(\cdot)$ is the median among the all agents $j\in\{1,...,n\}$.

Table~\ref{Tab:simulation_settings_fcn} and \ref{Tab:simulation_settings_fcn_searched} describe fixed and searched parameters for multi-agent simulations using ZI-, FCN-, and adFCN-Agents. Same as the proposed method, we conducted 300 trials of simulations for each baseline to obtain synthetic data and preprocess them using method proposed by \citet{ot_based_simulation_evaluation}. As a result of calibration, the selected parameters were $\lambda^c=1.50$, $\alpha=0.10$, $\tau=100$, and $\tau^{diff}=20$.

\begin{table}[htbp]
\centering
\caption{Fixed parameters for ZI-, FCN-, and adFCN-Agents.}
\label{Tab:simulation_settings_fcn}
\begin{tabular}{lll}
\toprule
\textbf{Parameter} & \textbf{Notation} & \textbf{Value} \\
\midrule
\# of agents     & $n$      & 200\\
\# of time steps & $T_{sim}$  & 2{,}110\\
Expected fundamental weight & $\lambda^f$ & 10.00\\
%Expected chartist weight    & $\lambda^c$ & 1.50\\
Expected noise weight       & $\lambda^n$ & 1.00\\
\makecell[l]{Standard deviation of\\ 
~~noise} & $\sigma^n$  & $0.01\times10^{-2}$\\
\makecell[l]{Mean reversion term of\\
~~fundamental price} & $\tau^f$ & $200$\\
Time window heterogeneity & $\tau^{diff}$ & $1.00$\\
Heterogeneity in risk aversion & $\alpha^{diff}$ & $1.00$\\
\bottomrule
\end{tabular}
\end{table}

\begin{table}[htbp]
\centering
\caption{Searched parameters for FCN- and adFCN-Agents.}
\label{Tab:simulation_settings_fcn_searched}
\begin{tabular}{lll}
\toprule
\textbf{Parameter} & \textbf{Notation} & \textbf{Candidate Values} \\
\midrule
Expected chartist weight    & $\lambda^c$ & $0.50,1.00,1.50,2.00$\\
Reference risk aversion level & $\alpha$ & $0.05,0.10,0.15,0.20$\\
Time window size & $\tau$   & $100,150,200$\\
Time window heterogeneity & $\tau^{diff}$ & $1,6,8,10,14,20$\\
\bottomrule
\end{tabular}
\end{table}

\section{\hashimoto{Data Description}}\label{App:data}

 We used FLEX-FULL historical tick data provided by the Japan Exchange Group~\citep{flex_full}. It includes order and execution series data, called tick data. By recording the mid price every minute, we obtained a one-minute bar price series with length $T_{len}=300$ per day. The data period was from January 5, 2015 to December 31, 2021. As described in Table~\ref{Tab:tickers}, we selected $22$ stocks from those with sufficiently high liquidity during the period, ensuring a diverse range of industries, and randomly divided them for calibrating and testing.

We used this dataset, a commercial data source, due to the absence of publicly available alternatives that meet the scientific requirements. Specifically, publicly available datasets lack sufficiently high-frequency resolution (e.g., one-minute bars), extended time spans, and coverage of a wide range of highly liquid stocks. FLEX-FULL provides comprehensive one-minute-level data over long periods across diverse, actively traded equities, which is essential for evaluating the realism and robustness of financial market simulations.

\begin{table}[htb]
    \centering
    \caption{Ticker codes of real data used for training deep generative models or calibrating ABMs, and testing.}
    \label{Tab:tickers}
    \begin{minipage}{0.49\hsize}
        \centering
        \subcaption{Train / Calibrate}
        \label{Tab:tickers4train}
        \begin{tabular}{cccc}
        \toprule
        3407 & 4188 & 4568 & 5020 \\
        6502 & 6758 & 7203 & 7550 \\
        8306 & 9202 & 9437 &      \\
        \bottomrule
        \end{tabular}
    \end{minipage}\hfill
    \begin{minipage}{0.49\hsize}
        \centering
        \subcaption{Test}
        \label{Tab:tickers4test}
        \begin{tabular}{cccc}
        \toprule
        2802 & 3382 & 4063 & 4452 \\
        4578 & 6501 & 7267 & 8001 \\
        8035 & 8058 & 9613 &\\
        \bottomrule
        \end{tabular}
    \end{minipage}
\end{table}

\end{document}